# HIERARCHICAL SPATIAL MODELS FOR PREDICTING TREE SPECIES ASSEMBLAGES ACROSS LARGE DOMAINS

By Andrew O. Finley[1], Sudipto Banerjee and Ronald E. McRoberts

*Michigan State University, University of Minnesota and USDA Forest Service*

Spatially explicit data layers of tree species assemblages, referred to as forest types or forest type groups, are a key component in large-scale assessments of forest sustainability, biodiversity, timber biomass, carbon sinks and forest health monitoring. This paper explores the utility of coupling georeferenced national forest inventory (NFI) data with readily available and spatially complete environmental predictor variables through spatially-varying multinomial logistic regression models to predict forest type groups across large forested landscapes. These models exploit underlying spatial associations within the NFI plot array and the spatially-varying impact of predictor variables to improve the accuracy of forest type group predictions. The richness of these models incurs onerous computational burdens and we discuss dimension reducing spatial processes that retain the richness in modeling. We illustrate using NFI data from Michigan, USA, where we provide a comprehensive analysis of this large study area and demonstrate improved prediction with associated measures of uncertainty.

**1. Introduction.** Forest type is a classification of forestland based on the plurality of species of all live trees that contribute to stocking. Stocking, in turn, is a measure of actual forest stand density relative to the density considered optimal for a desired purpose, such as site occupancy or volume growth [Stage (1969)]. A forest type group (FTG) is an assemblage of forest cover types that share closely associated forest species or site requirements.

Forest area by tree species composition classes, such as forest types and FTGs, has received increased attention in recent years as an indicator of

Received October 2008; revised April 2009.

[1]Supported by NSF Grant DMS-07-06870 and the USDA Forest Service FIA and FHTET programs.

forest sustainability and biodiversity. The Ministerial Conference on the Protection of Forests in Europe (MCPFE 2008: http://www.mcpfe.org) includes area by forest type as an indicator for a criterion related to maintaining forest resources, and the Montréal Process (Montréal Process 2005: www.rinya.maff.go.jp/mpci) includes the same indicator for criterion related to maintaining ecosystem biodiversity and forest productivity. Action E43 (Harmonization of the national forest inventories of Europe) (COST E43 2007: www.metla.fi/eu/cost/e43) of the European program of Cooperation in the field of Scientific and Technical Research has selected forest type as a core variable for biodiversity assessments. Commercial enterprises in North American, Mediterranean, central European and Nordic countries also rely on spatially explicit regional and national forest assessments by type to support decisions regarding establishment or expansion of facilities (e.g., paper mills) and for long-term forecasts of wood fiber supplies for burgeoning energy bioeconomies.

National forest inventories (NFI) conducted in North America, Europe and elsewhere are the most important sources of comprehensive information for assessing FTGs for large geographic domains. Because complete enumerative inventories are prohibitively expensive, NFIs sample populations of interest and report plot-based estimates of forest resources. For valid sampling designs and corresponding estimators, these plot-based approaches produce asymptotically unbiased estimates of area by forest types and FTGs. However, these approaches are unable to depict spatial distributions of forest attributes and do not easily incorporate ancillary variables or complex spatial dependence structures to improve the accuracy and precision of parameter estimates and/or prediction. Therefore, model-based approaches to mapping are attracting greater interest. Natural resource mapping initiatives typically entail constructing statistical models of relationships between land cover attributes and variables including soil, climatic, topographic and satellite image spectral variables. These models are then used to produce digital data layers of small-area spatially explicit prediction across large domains, which ultimately support the end-user analyses described above.

Spatial process models [e.g., Cressie (1993); Stein (1999)] to analyze NFI data have, hitherto, focused largely upon continuous outcomes such as prediction of biomass [e.g., Finley et al. (2008a)]. A classic paper by Diggle, Tawn and Moyeed (1998) discussed the use of spatial process models for non-Gaussian data within the framework of generalized linear models. Heagerty and Lele (1998) considered a composite likelihood approach to binary spatial regression. Recently, Finley, Banerjee and McRoberts (2008b) explored a spatial logistic regression model to predict forested areas. Our data here concerns categorical (seven FTG's) rather than binary outcomes. Furthermore, realizations of a given FTG will exhibit different composition and proportion of species. For instance, two oak dominated plots assigned to the oak/hickory FTG might have different water or temperature requirements due to the type of oak species present and proximity along environmental gradients (e.g., springtime water availability or minimum annual temperature). We, therefore, find it attractive to allow the regression coefficients to vary by location, envisioning a spatial surface associated with each coefficient. For instance, we could model the spatial surface for the coefficient parametrically—using perhaps a polynomial surface function. Such specifications are often too arbitrary and may lead to a range of surfaces too inflexible for our purposes. In related work, Fahrmeir and Lang (2001) and Kneib and Fahrmeir (2006) consider semiparametric regression with splines and Markov random fields to model spatial effects. Modeling multiple regression coefficient processes jointly, however, would require multivariate specifications of splines that may be awkward. Instead, we treat each regression coefficient surface as a realization from a continuous spatial process. A multivariate spatial process is arguably more natural and at least as flexible here.

We use spatially-varying multinomial logistic regression models to exploit, more fully, the spatial proximity of the NFI plot array and the potentially spatially-varying impact of the predictor variables on the response to improve the accuracy and precision of FTG prediction at locations where we have observed predictors but not inventory plots. We also encounter a large number of locations that make full inference computationally onerous. Modeling large spatial datasets has received much recent attention (see Section 4), but many of the existing approaches are unsuitable for our coefficient processes. We discuss how a low-rank spatial process can be adapted to model the regression coefficients and achieve computational feasibility.

While most of the models we formulate can possibly be estimated using maximum likelihood or variants thereof, we adopt a Bayesian approach [e.g., Gelfand et al. (2003)]. This is attractive, as it offers exact inference for the random spatial coefficients, and that too with non-Gaussian data, by delivering an entire posterior distribution at both observed and unobserved locations. Spatial interpolation for processes that are neither observed nor arise as residuals appears inaccessible with classical likelihood-based methods. On the other hand, Bayesian model fitting involves rather specialized Markov chain Monte Carlo (MCMC) methods [see, e.g., Robert and Casella (2005)] that raise concerns about computational expense and reproducibility of inference. These concerns have, however, started to wane with the delivery of relatively simpler `R` packages (www.r-project.org), including `mcmc`, `MCMCpack`, `geoRglm` and `spBayes`, that help automate such methods and diagnose convergence.

While our primary contribution here lies in the novel application, we also offer several methodological advancements. As mentioned earlier, we extend existing spatial logistic models to spatially-varying multinomial regression models. This involves multivariate spatial processes (one for each regression coefficient) that we wish to model jointly. This approach is similar to Gelfand et al. (2003), but, unlike there, we allow each coefficient process to have its own spatial correlation structure. We achieve this using linear transformations of independent processes. This idea has been used elsewhere [e.g., Finley et al. (2008a)] to model multivariate continuous outcomes, but not completely unobserved coefficient processes as we attempt here.

The remainder of the article proceeds as follows. Section 2 provides an overview of the NFI data and study area used to illustrate our proposed methods. These descriptions are followed by a brief preliminary analysis, the results of which help motivate the models and methods presented in Sections 3 and 4. The full analysis of these NFI data using the proposed models is detailed in Section 5. In Section 6 we present and discuss the results of this analysis. Finally, Section 7 concludes the paper with a brief summary and description of future direction and work.

**2. Data.** The Forest Inventory and Analysis (FIA) program of the U.S. Forest Service conducts the NFI of the USA. The program has established field plot centers in permanent locations using a sampling design that produces an equal probability sample [Bechtold and Patterson (2005)]. The sampling design is based on a tessellation of the USA into approximately 2400 ha hexagons and features a permanent plot at a randomly selected location within each hexagon. The State of Michigan in which the study area is located provided additional funding to triple the sampling intensity to approximately one plot per 800 ha. Each plot consists of four 7.32 m radius circular subplots for a total area of 672 $m^2$. The subplots are configured as a central subplot and three peripheral subplots with centers located at 36.58 m and azimuths of $0°$, $120°$ and $240°$ from the center of the central subplot. In general, locations of forested or previously forested plots are determined using global positioning system receivers, whereas locations of nonforested plots are verified using aerial imagery and digitization methods.

Field crews observe species and measure diameter at-breast-height (dbh) (1.37 m) and height for all trees with dbh of 12.7 cm or greater and assign forested portions of subplots to forest types based on visual assessments. Forest types are also assigned to forest portions of subplots using algorithms based on measures of density and stocking calculated from species observations and diameter and height measurements of all plot trees. To mitigate the effects of plot location errors and eliminate the difficulties associated with plots to which multiple FTGs were assigned, we use only the central subplot to which a single FTG had been assigned.

2.1. *Study area.* The study area is the 79,094.17 $km^2$ forested land of Michigan. Due to proximity to the Great Lakes and numerous episodes of glaciation, this region exhibits a host of climate and soil characteristics that have produced diverse forest communities. Precipitation and temperature extremes are well-known factors in determining spatial patterns of forest species composition in the Great Lakes states [Albert (1995)]. Therefore, the predictor variables we consider include a set of long-term climate data and a soil drainage index. We obtained raster data layers of mean annual precipitation (PRECIP), temperature minimum (TMIN) and temperature maximum (TMAX) over the period 1971–2000, and average annual snowfall (SNOW) over the period of 1961–1990. These data were generated by the PRISM climate mapping project [Daly et al. (2000)]. Recently, Henne, Hu and Cleland (2007) showed that from a suite of long-term monthly climate and soil composition variables, lake-effect snowfall abundance contributes the most to explaining spatial variation in mesic tree species (i.e., species such as sugar maple or beech that are found on sites with moderate soil moisture) within the Lake State region. These authors also suggest that spring lake-effect snow provides moisture to course-textured xeric soil, allowing mesic forest types to become established on these otherwise droughty soils. The long-term climate patterns and soil characteristics affect tree species survival and influence the assemblage of species found on a given site [Host et al. (1988)].

The soil drainage index (DI) raster data layer was formulated to mimic the quantity of water present in a soil and is available to plants under normal, long-term climatic conditions, including water under saturated and unsaturated conditions [Schaetzl (1986)]. The DI variable ranges from 0 for the driest soils to 99 for open water. The PRISM data layers were originally generated at $\sim$0.8$\times$0.8 km pixel resolution, but were resampled to match the 30$\times$30 m resolution of the DI data layer. We make this simplifying assumption because the disparity between the data layer resolution is small compared to the distance between FIA plot observations. Finally, all predictor data layers were reprojected to share the projection of the georeferenced forest inventory plots (Figure 1). Here we see strong spatial dependence among the predictor variables, for instance, north to south gradients in precipitation and temperature extremes and gradients in mean temperature minimum and mean snow depth that are perpendicular to the shorelines.

Figure 2(a) illustrates the georeferenced forest inventory data consisting of 5180 forested FIA plots measured between 1999 and 2006 that meet our inclusion criterion. For this analysis, the FTGs of interest and associated relative percent observed across the inventory plots are white/red/jack pine (10.89%), spruce/fir (14.56%), oak/pine (2.86%), oak/hickory (12.92%), elm/ash/cottonwood (6.60)%, aspen/birch (16.99%), and maple/beech/birch (35.19%). The left column in Figures 3 and 4 offers interpolated surfaces of the FTGs. Note that interpolation is over binary values that indicate the presence and absence of the given FTG. The surfaces show strong spatial patterns in FTG range and extent.

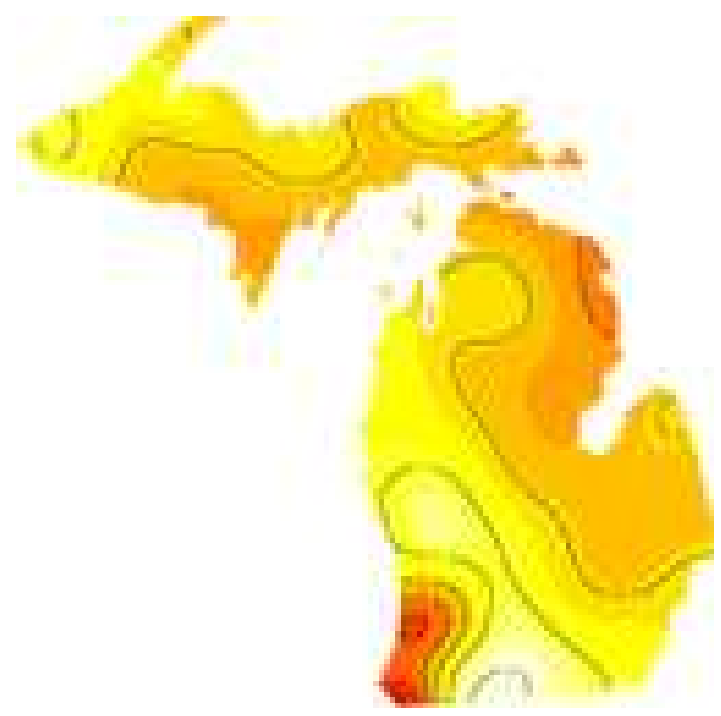

(a) Mean annual precipitation

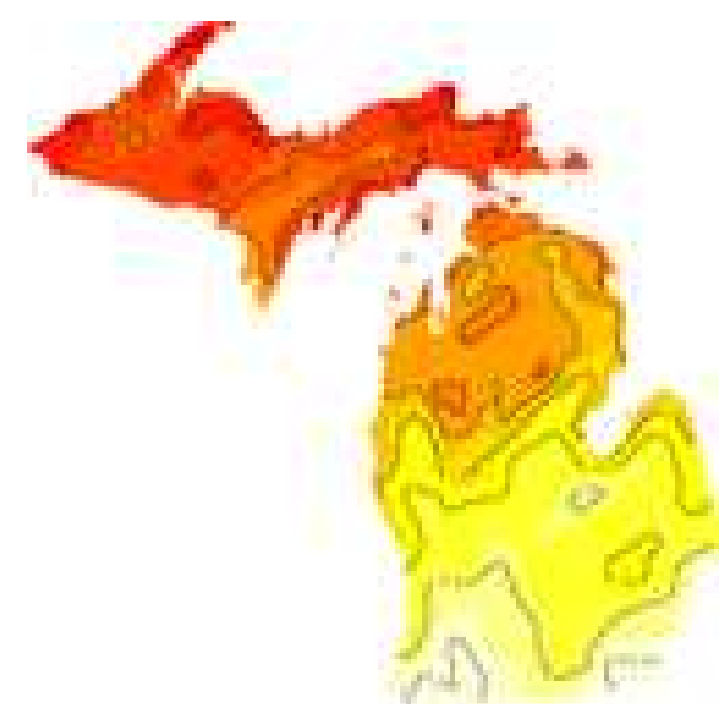

(b) Mean annual temperature max

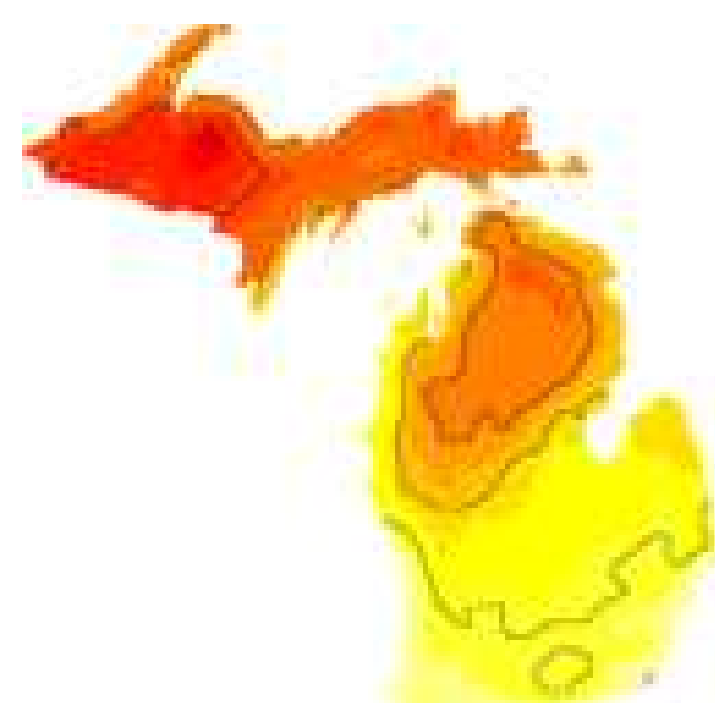

(c) Mean annual temperature min

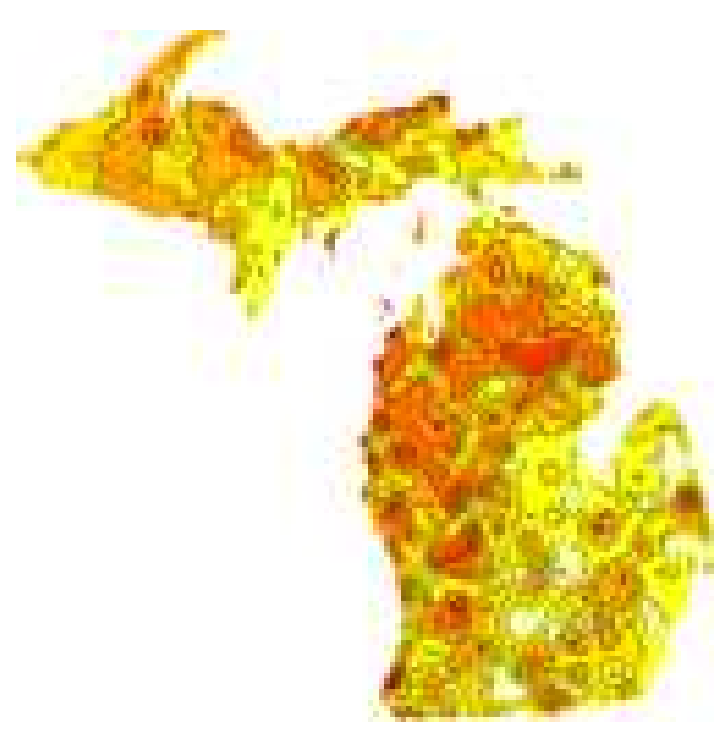

(d) Soil drainage index

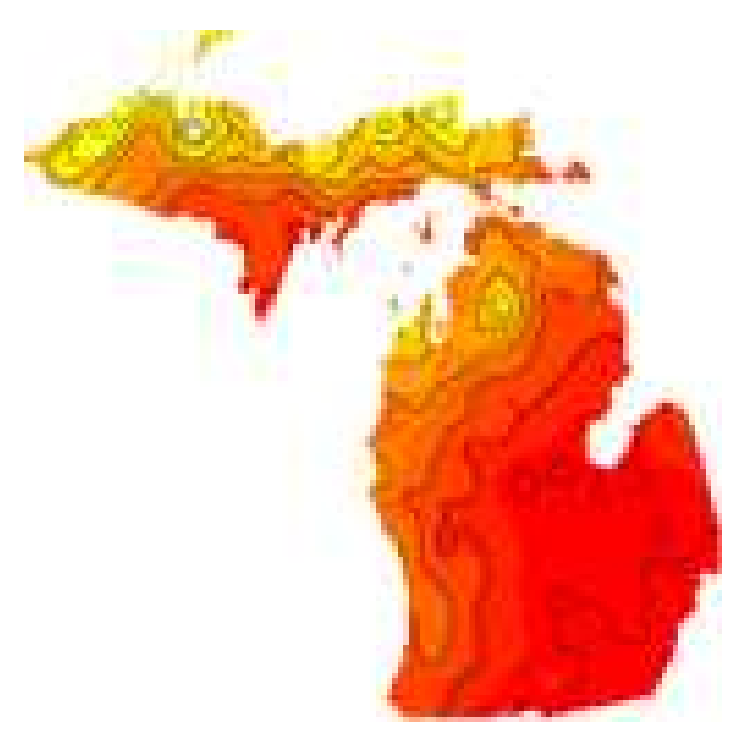

(e) Mean annual snow depth

FIG. 1. *Surfaces of the mean zero and unit variance standardized predictor variables resampled to a 30×30 meter resolution across the domain. Lighter colors correspond to higher values.*

2.2. *Preliminary analyses.* We index the georeferenced FIA plots as $\mathcal{S} = \{\mathbf{s}_1, \ldots, \mathbf{s}_n\}$, where $\mathbf{s}$ is a coordinate vector (e.g., longitude and latitude), and use a binary outcome $Y(\mathbf{s}_i) = 1$ or 0 to indicate the presence or ab-sence of a given FTG for each location in $\mathcal{S}$ which depends, in part, upon predictors/regressors $\mathbf{x}(\mathbf{s})$ for a generic plot $\mathbf{s}$. Given this setting, a logistic regression model is the natural choice for relating the outcome to its predictors. It is, however, unrealistic to assume that the model parameter values associated with the predictor variables are constant across the study area. As noted in Section 1, FTGs are not *true* categorical variables because they are based on somewhat artificial levels of continuous variables, such as stocking and proportions of species. For instance, plots labeled aspen/birch will exhibit a continuum of aspen and birch tree species composition (e.g., from pure aspen to pure birch). Although we expect the assemblage of tree species within a given FTG to generally occur together due to shared environmental requirements, the observed realization of the FTG across the landscape will exhibit some level of heterogeneity in species composition. As described above, this is a large domain that contains several broad climate, soil origin and topographic gradients. Therefore, within a given FTG, we expect regionalized inter-species and intra-species (e.g., genetic adaptation) response to predictor variables.

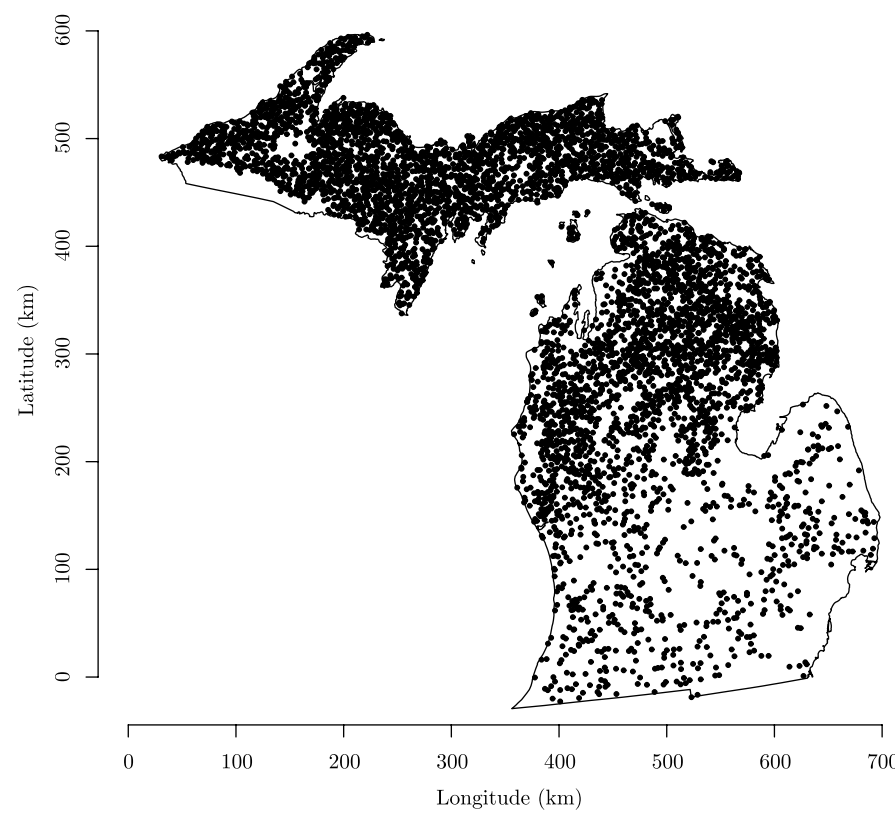

(a) FIA plots

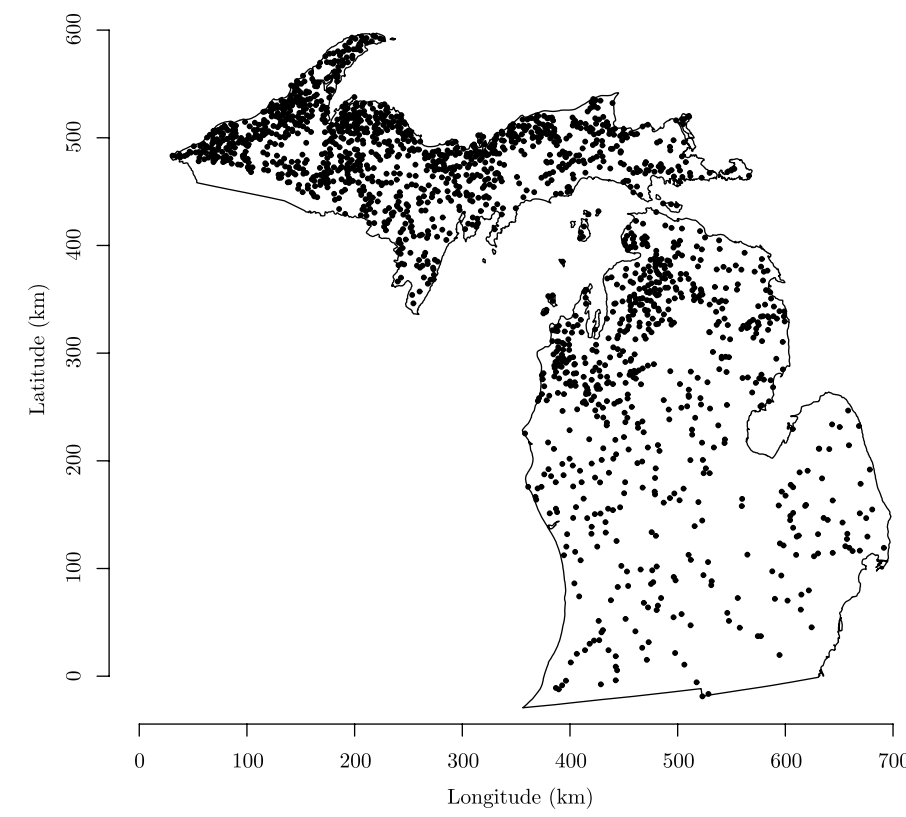

(b) Maple/beech/birch FTG plots

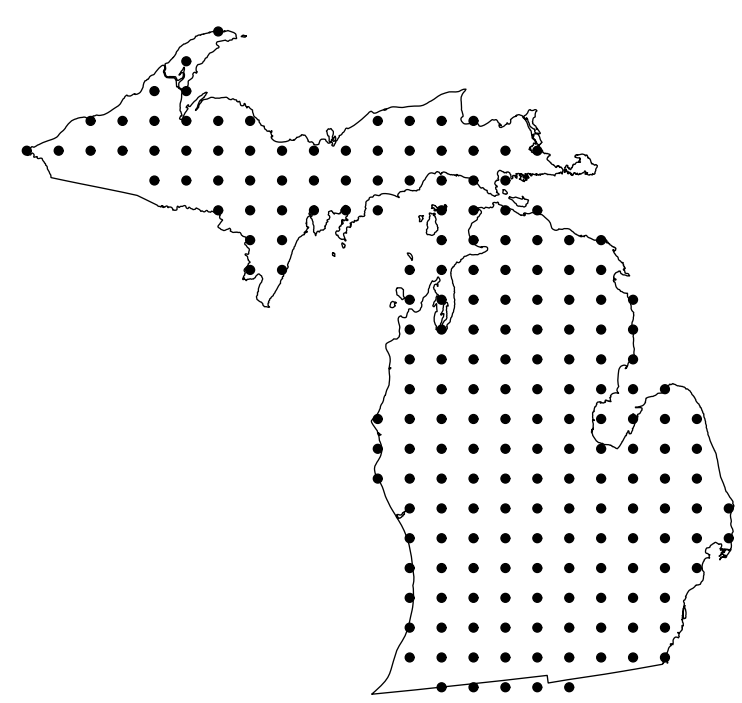

(c) Predictive process knots

FIG. 2. *Forest inventory plot locations, plots locations belonging to the maple/beech/birch FTG, and candidate predictive process knot configuration.*

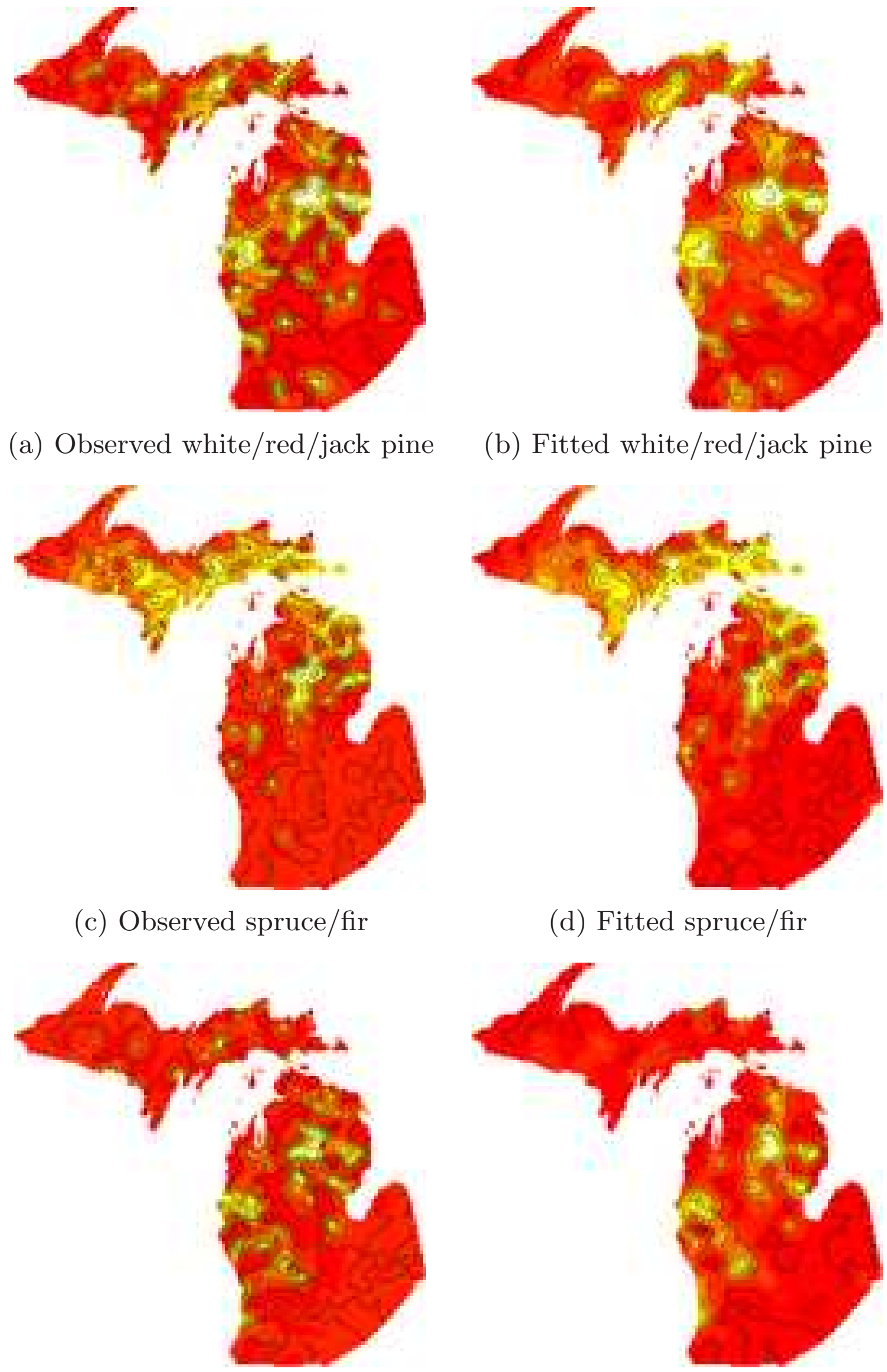

(a) Observed white/red/jack pine (b) Fitted white/red/jack pine

(c) Observed spruce/fir (d) Fitted spruce/fir

(e) Observed oak/pine (f) Fitted oak/pine

FIG. 3. *Interpolated surfaces of the observed and 200 knot spatially-varying coefficients model's fitted FTGs. For the observed surfaces the interpolation is over the 0 or 1 binary variable of FTG presence/absence and the fitted surfaces are over the probability of FTG occurrence. Lighter colors correspond to higher values. The remaining FTG and associated fitted values are offered in Figure 4.*

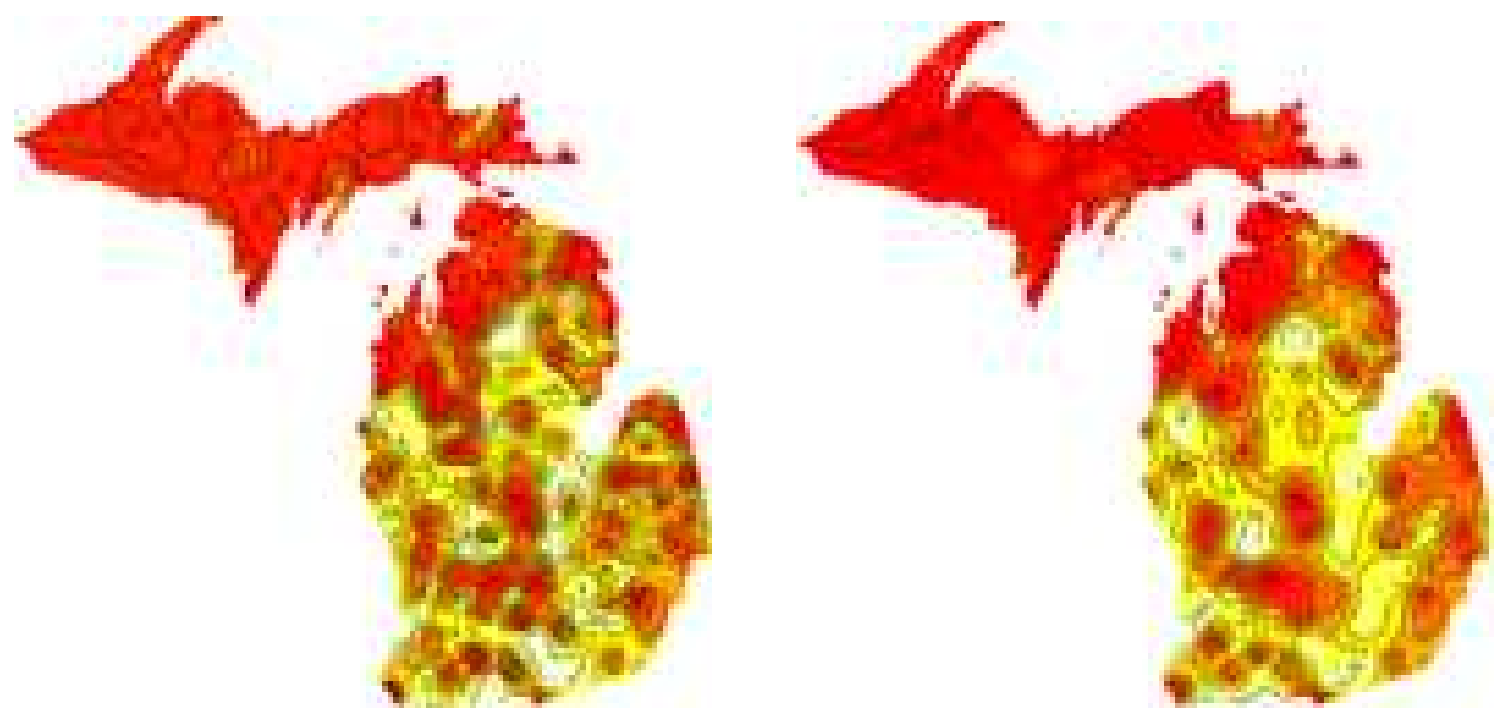

(a) Observed oak/hickory (b) Fitted oak/hickory

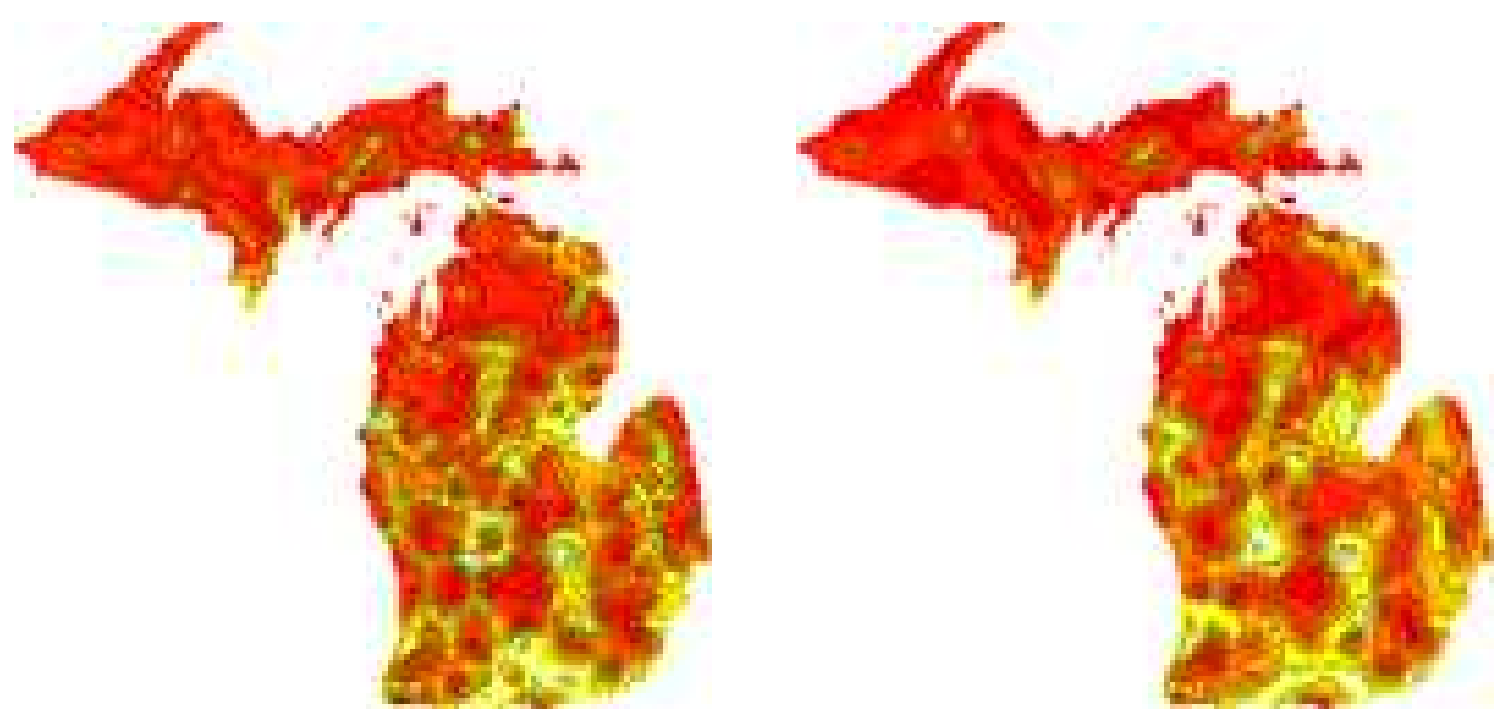

(c) Observed elm/ash/cottonwood (d) Fitted elm/ash/cottonwood

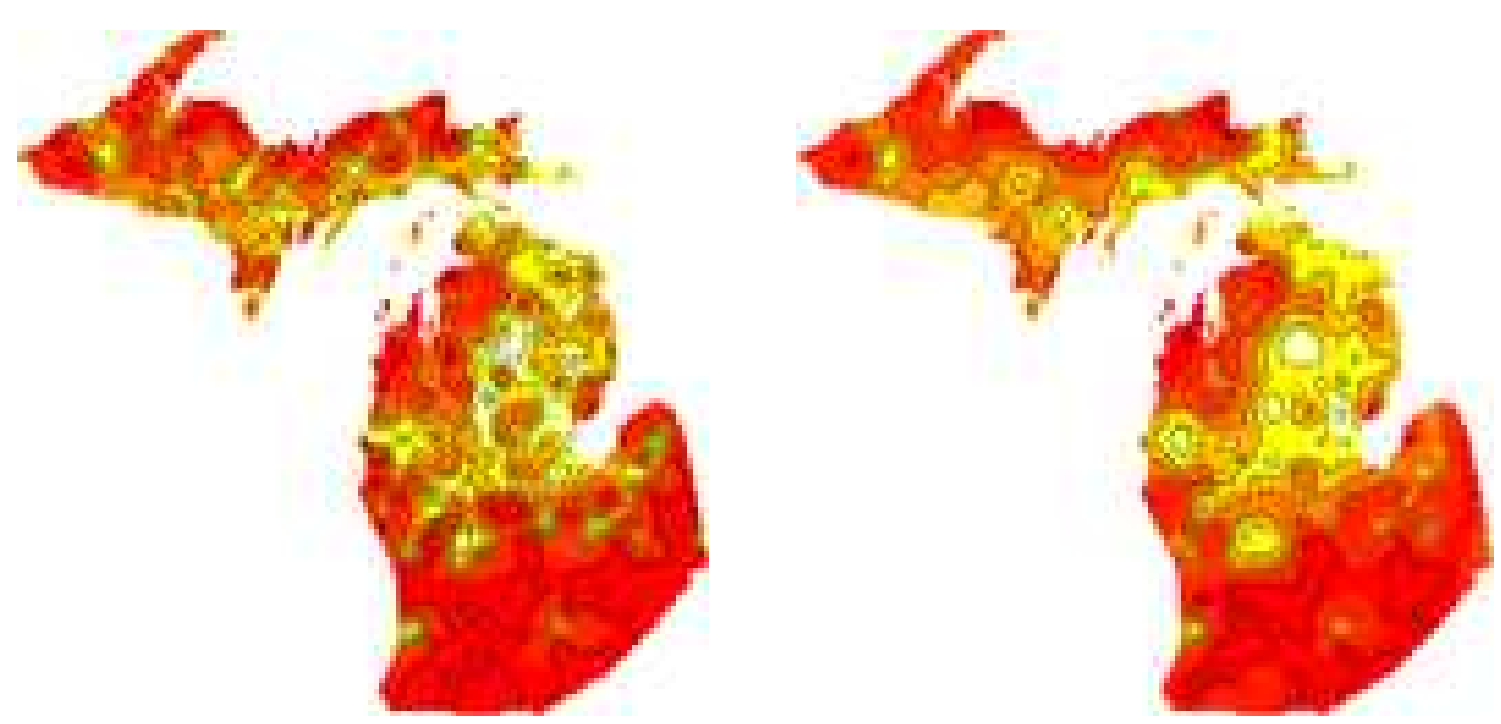

(e) Observed aspen/birch (f) Fitted aspen/birch

FIG. 4. *Interpolated surfaces of the observed and 200 knot spatially-varying coefficients model's fitted FTGs. For the observed surfaces the interpolation is over the 0 or 1 binary variable of FTG presence/absence and the fitted surfaces are over the probability of FTG occurrence. Lighter colors correspond to higher values.*

We pose a simple-minded analysis to explore variation or, more specifically, the nonstationarity of predictor variables' parameter estimates for the presence/absence of a single FTG. We tessellate the domain such that each tessera holds some minimum number of FIA plots, say, 10, then include a tessera random effect on each predictor variable. For this analysis the domain was covered with 68 hexagon tessera indexed by $k = 1, \ldots, 68$. For a given generic location $\mathbf{s}$ the logistic regression model is given by

$$p(\mathbf{s}) = \frac{\exp(\mathbf{x}(\mathbf{s})'\hat{\boldsymbol{\beta}}(\mathbf{s}))}{1 + \exp(\mathbf{x}(\mathbf{s})'\hat{\boldsymbol{\beta}}(\mathbf{s}))}, \tag{2.1}$$

where $p(\mathbf{s})$ is the probability that location, or inventory plot, $\mathbf{s}$ belongs to the given FTG, $\mathbf{x}(\mathbf{s})$ is a $p \times 1$ vector that holds an intercept and our set of regressors/predictors which include long-term climate and soil variables (detailed in Section 2.1), and $\hat{\beta}(\mathbf{s})$ are tessera varying coefficients. We can decompose the adaptive $\hat{\beta}(\mathbf{s}) = \boldsymbol{\beta} + \mathbf{u}_k$, where $\mathbf{u}_k$ is the $k$th tessera's vector of random effects (i.e., the tessera within which the given $\mathbf{s}$ falls). We assume that $\mathbf{u}_k \stackrel{\text{ind}}{\sim} MVN(\mathbf{0}, \Psi)$ is a multivariate normal distribution and, for simplicity, $\Psi = \mathrm{diag}[\tau_i^2]_{1=i}^p$. This model was fit for each FTG. We assume that the $\beta$'s have *flat* prior distributions and the $\tau^2$'s follow an inverse-Gamma ($IG$) with hyperparameters $IG(2, 1)$. Note, with a shape value of 2, the $IG$ distribution has infinite variance and is centered on the scale value, which in this case is 1. We experimented with a range of scale values but saw negligible change in the final parameter estimates. Three MCMC chains of 75,000 iterations were run for each model. Then posterior inference was based on $3 \times 50{,}000 = 150{,}000$ post burn-in samples.

For brevity, Table 1 shows only parameter estimates for white/red/jack pine and spruce/fir FTGs. Several predictor variables' parameter estimates for these and the other FTGs are significant at the 0.05 level. Also, the relative magnitude of the variance terms suggests that we should consider models that explicitly accommodate spatial nonstationarity of $\boldsymbol{\beta}$. This idea is further supported by the presence of clustering seen in maps of $\beta$ specific random effects; see, for example, maps for red/white/jack pine FTG in Figure 5. Similar clustering patterns are seen when mapping the $\mathbf{u}$ for the other FTGs as well. Given dependence structures capable of modeling this potential spatial nonstationarity, we will improve model fit and predict FTG with greater accuracy and precision across the study area.

## 3. Models.

3.1. *Spatially-varying multinomial logistic regression models.* As described in Section 1, using the tree species composition and relative stocking, FTGs are assigned to inventory plots. Our initial interest lies in modeling the probability of a given plot belonging to a specific FTG. We can obtain these probabilities by using $J-1$ baseline-category logistic regressions models [e.g., Agresti (2002), Chapter 7]. For a generic location $\mathbf{s}$, we consider $J-1$ binary outcome variables such that $Y_j(\mathbf{s}) = 1$ if the given plot belongs to the $j$th FTG, $j = 1, \ldots, J-1$, and 0 if it belongs to the $J$th (baseline) FTG. The spatially-varying baseline-category logistic regressions yield

$$\pi_j(\mathbf{s}) = \frac{\exp(\mathbf{x}_j(\mathbf{s})'\tilde{\boldsymbol{\beta}}_j(\mathbf{s}))}{1 + \sum_{k=1}^{J-1} \exp(\mathbf{x}_k(\mathbf{s})'\tilde{\boldsymbol{\beta}}_k(\mathbf{s}))}, \tag{3.1}$$

where $\pi_j(\mathbf{s})$ is the probability that location $\mathbf{s}$ belongs to the $j$th FTG, $\mathbf{x}_j(\mathbf{s})$ was defined previously, and $\tilde{\boldsymbol{\beta}}(\mathbf{s})$ denotes spatially-varying coefficients. Then, given the $J-1$ models, the set of probabilities for $\mathbf{s}$ are completed by calculating $\pi_J(\mathbf{s}) = 1 - \sum_{j=1}^{J-1} \pi_j(\mathbf{s})$. The baseline category is customarily set to the most commonly occurring and spatially pervasive FTG.

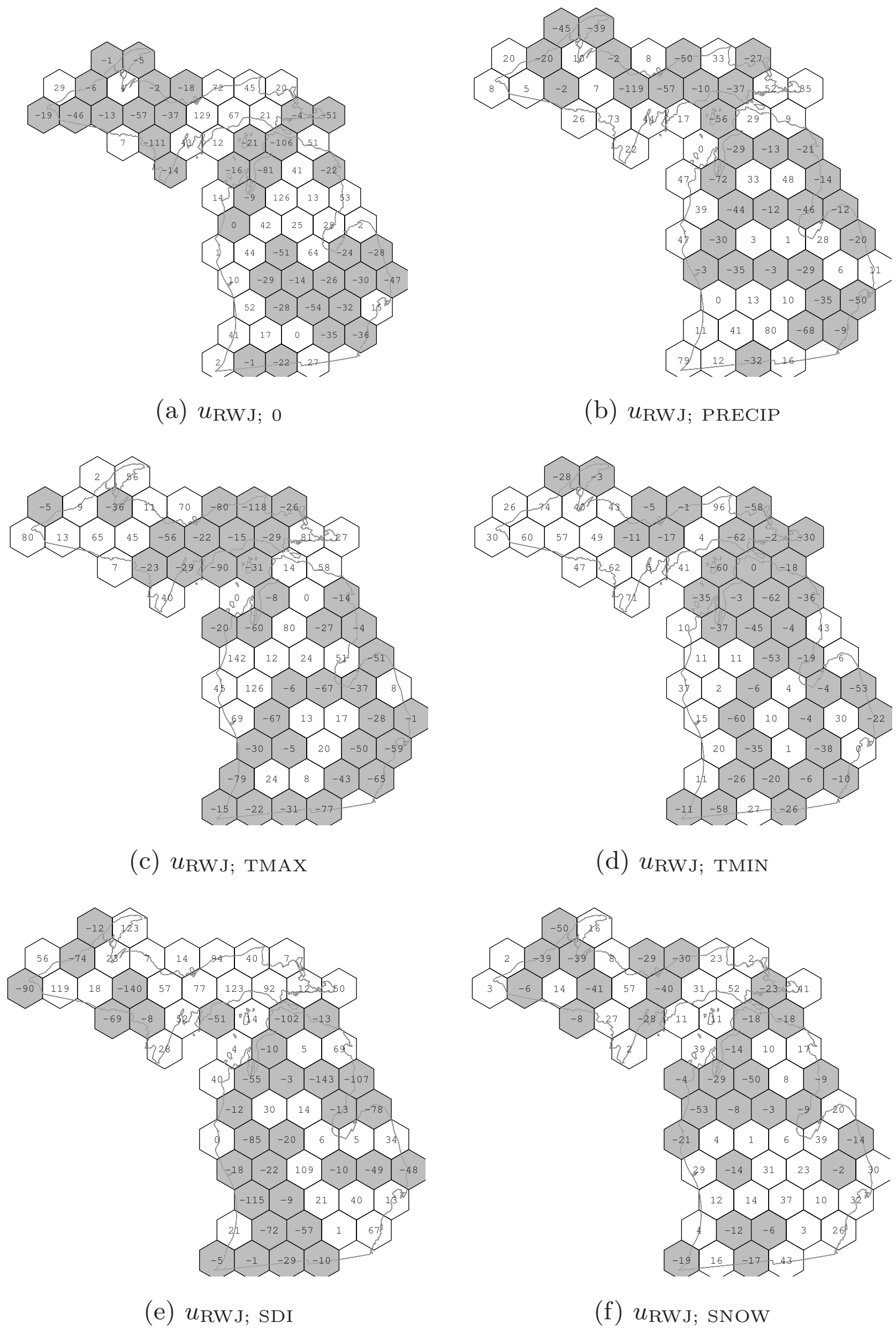

(a) $u_{\mathrm{RWJ};\,0}$ (b) $u_{\mathrm{RWJ};\,\mathrm{PRECIP}}$

(c) $u_{\mathrm{RWJ};\,\mathrm{TMAX}}$ (d) $u_{\mathrm{RWJ};\,\mathrm{TMIN}}$

(e) $u_{\mathrm{RWJ};\,\mathrm{SDI}}$ (f) $u_{\mathrm{RWJ};\,\mathrm{SNOW}}$

FIG. 5. *Hexagon specific intercept and predictor variable random effects for the red/white/jack pine (RWJ) FTG (values scaled by 100).*

Table 1
*Parameter credible intervals for the red/white/jack pine (RWJ) and spruce/fir (SF) FTGs for the hexagon specific random effects models*

| **Parameter** | 2.5% | 50% | 97.5% | **Parameter** | 2.5% | 50% | 97.5% |
|---|---|---|---|---|---|---|---|
| $\beta_{\text{RWJ; 0}}$ | −2.06 | −1.76 | −1.50 | $\tau^2_{\text{RWJ; 0}}$ | 0.27 | 0.46 | 0.75 |
| $\beta_{\text{RWJ; PRECIP}}$ | −0.55 | −0.19 | 0.18 | $\tau^2_{\text{RWJ; PRECIP}}$ | 0.24 | 0.41 | 0.68 |
| $\beta_{\text{RWJ; TMAX}}$ | −0.19 | 0.12 | 0.44 | $\tau^2_{\text{RWJ; TMAX}}$ | 0.32 | 0.52 | 0.94 |
| $\beta_{\text{RWJ; TMIN}}$ | −0.75 | −0.43 | 0.03 | $\tau^2_{\text{RWJ; TMIN}}$ | 0.14 | 0.34 | 0.86 |
| $\beta_{\text{RWJ; SDI}}$ | −1.02 | −0.76 | −0.47 | $\tau^2_{\text{RWJ; SDI}}$ | 0.45 | 0.66 | 0.98 |
| $\beta_{\text{RWJ; SNOW}}$ | −0.69 | −0.40 | −0.02 | $\tau^2_{\text{RWJ; SNOW}}$ | 0.17 | 0.26 | 0.42 |
| $\beta_{\text{SF; 0}}$ | −2.47 | −2.15 | −1.89 | $\tau^2_{\text{SF; 0}}$ | 0.22 | 0.41 | 0.63 |
| $\beta_{\text{SF; PRECIP}}$ | −0.88 | −0.53 | −0.23 | $\tau^2_{\text{SF; PRECIP}}$ | 0.19 | 0.31 | 0.50 |
| $\beta_{\text{SF; TMAX}}$ | −1.27 | −0.85 | −0.45 | $\tau^2_{\text{SF; TMAX}}$ | 0.22 | 0.41 | 0.74 |
| $\beta_{\text{SF; TMIN}}$ | −1.31 | −0.84 | −0.49 | $\tau^2_{\text{SF; TMIN}}$ | 0.19 | 0.33 | 0.60 |
| $\beta_{\text{SF; SDI}}$ | 1.57 | 1.79 | 2.10 | $\tau^2_{\text{SF; SDI}}$ | 0.16 | 0.30 | 0.56 |
| $\beta_{\text{SF; SNOW}}$ | −0.74 | −0.40 | −0.02 | $\tau^2_{\text{SF; SNOW}}$ | 0.15 | 0.26 | 0.42 |

We decompose the coefficients as $\tilde{\boldsymbol{\beta}}_j(\mathbf{s}) = \boldsymbol{\beta}_j + \mathbf{w}_j(\mathbf{s})$, where $\boldsymbol{\beta}_j$'s represent the nonspatial regression coefficients, as in customary logistic regression, while $\mathbf{w}_j(\mathbf{s})$ is a *multivariate* spatial process. More generally, we can write the regression component as $\mathbf{x}_j(\mathbf{s})'\boldsymbol{\beta} + \mathbf{z}_j(\mathbf{s})'\mathbf{w}_j(\mathbf{s})$, where $\mathbf{z}_j(\mathbf{s}) = \mathbf{x}_j(\mathbf{s})$ for a fully spatially-varying coefficients candidate model, or could be a subvector of $\mathbf{x}_j(\mathbf{s})$ representing those predictors whose impact is posited to vary spatially; for instance, a spatially varying intercept model will correspond to $\mathbf{z}_j(\mathbf{s})$ being equal to 1. In principle, we could further hypothesize dependence across the $j$ categories. This, however, is not intuitive in our context and further complicates the modeling without clear gains. Hence, we assume $\mathbf{w}_j(\mathbf{s})$'s are independent across $j$, but each $\mathbf{w}_j(\mathbf{s})$ is a vector of correlated processes.

3.2. *Multivariate process models.* Multivariate spatial processes, such as each $\mathbf{w}_j(\mathbf{s})$, are completely characterized by their mean and a cross-covariance (matrix) function (suppressing the suffix on categories), $\mathbf{C_w}(\mathbf{s}_1, \mathbf{s}_2; \boldsymbol{\theta}) = \text{cov}\{\mathbf{w}(\mathbf{s}_1), \mathbf{w}(\mathbf{s}_2)\}$. Valid constructions using convolutions of kernels or correlation functions are possible [Ver Hoef and Barry (1998); Gaspari and Cohn (1999)]. An attractive, easily interpretable and flexible approach develops versions of the linear model of coregionalization (LMC) as in, for example, Grzebyk and Wackernagel (1994), Wackernagel (2006), Schmidt and Gelfand (2003) or Gelfand et al. (2004). See, also, Reich and Fuentes (2007) for a Bayesian nonparametric adaptation.

In the LMC approach we let $\mathbf{w}(\mathbf{s}) = \mathbf{A}(\mathbf{s})\mathbf{v}(\mathbf{s})$ be a spatial linear transformation, where $\mathbf{v}(\mathbf{s}) = (v_1(\mathbf{s}), \dots, v_p(\mathbf{s}))'$ and each $v_i(\mathbf{s})$ is an independent spatial process with unit variance and correlation function $\rho_i(\mathbf{s}_1, \mathbf{s}_2; \boldsymbol{\theta}_i)$. Thus, $\mathbf{v}(\mathbf{s})$ has a diagonal cross-covariance matrix $\mathbf{C_v}(\mathbf{s}_1, \mathbf{s}_2)$ with $i$th diagonal element as $\rho_i(\mathbf{s}_1, \mathbf{s}_2; \boldsymbol{\theta}_i)$ yielding a valid nonstationary cross-covariance $\mathbf{C_w}(\mathbf{s}_1, \mathbf{s}_2) = \mathbf{A}(\mathbf{s}_1)\mathbf{C_v}(\mathbf{s}_1, \mathbf{s}_2)\mathbf{A}(\mathbf{s}_2)'$ for $\mathbf{w}(\mathbf{s})$. A flexible choice for each $\rho_i(\mathbf{s}_1, \mathbf{s}_2; \boldsymbol{\theta}_i)$ is the Matérn correlation function, which allows control of spatial range, $\phi$, and smoothness, $\nu$ [Stein (1999)] and is given by

$$\rho(\mathbf{s}, \mathbf{s}'; \phi, \nu) = \frac{1}{2^{\nu-1}\Gamma(\nu)}(\|\mathbf{s} - \mathbf{s}'\|\phi)^\nu \mathcal{K}_\nu(\|\mathbf{s} - \mathbf{s}'\|; \phi); \qquad \phi > 0, \nu > 0. \tag{3.2}$$

In general, $\mathbf{w}(\mathbf{s})$ is nonstationary even when $\mathbf{v}(\mathbf{s})$ is stationary. When $\mathbf{A}(\mathbf{s}) = \mathbf{A}$, $\mathbf{w}(\mathbf{s})$ inherits stationarity from $\mathbf{v}(\mathbf{s})$: $\mathbf{C_w}(\mathbf{s}_1 - \mathbf{s}_2) = \mathbf{A}\mathbf{C_v}(\mathbf{s}_1 - \mathbf{s}_2)\mathbf{A}'$. Since $\mathbf{C_w}(\mathbf{s}_1, \mathbf{s}_2) = \mathbf{A}(\mathbf{s}_1)\mathbf{A}(\mathbf{s}_2)'$, one can assume that $\mathbf{A}(\mathbf{s}) = \mathbf{C_w}^{1/2}(\mathbf{s}, \mathbf{s})$ is a lower-triangular square-root; the one-to-one correspondence between the elements of $\mathbf{A}(\mathbf{s})$ and $\mathbf{C_w}(\mathbf{s}, \mathbf{s})$ is well known [e.g., Harville (1997), page 229]. Stationarity implies $\mathbf{A}(\mathbf{s}) = \mathbf{A}$ and that $\mathbf{C_w}(\mathbf{0}) = \mathbf{A}\mathbf{A}'$. Here we could either assign a prior, for example, inverse Wishart, to $\mathbf{A}\mathbf{A}'$ or could further parameterize it in terms of eigenvalues and the Givens angles which are themselves assigned hyperpriors [Daniels and Kass (1999)].

Once a valid cross-covariance function is specified for a multivariate Gaussian process, the realizations of $\mathbf{w}(\mathbf{s})$ over the set of observed locations $\mathcal{S}$ is given by $N(\mathbf{0}, \Sigma_\mathbf{w}(\boldsymbol{\theta}))$, where $\Sigma_\mathbf{w}(\boldsymbol{\theta})$ is an $np \times np$ block matrix whose $(i, j)$th block is the $p \times p$ cross-covariance $\mathbf{C_w}(\mathbf{s}_i, \mathbf{s}_j; \boldsymbol{\theta})$, $i, j = 1, \dots, n$. Without further specifications, estimating (3.1) will involve computing the inverse and determinant of the dense matrix $\Sigma_\mathbf{w}(\boldsymbol{\theta})$. Such computations invoke solvers or factorizations of complexity $O(n^3p^3)$, not once but iteratively, to produce estimates of $\boldsymbol{\theta}$. With large $n$, this is computationally infeasible.

**4. Multivariate process models for large datasets.** Modeling large spatial datasets observed over irregular locations has received much attention in the recent past. One approach employs spectral approximations to the likelihood [e.g., Stein (1999) and references therein; Fuentes (2002)], thereby avoiding large matrix computations [Paciorek (2007); Fuentes (2007)]. This works best assuming some form of stationarity and does not easily adapt to multivariate coefficient processes. Stein, Chi and Welty (2004) improve upon an idea of Vecchia (1988) [also, see Jones and Zhang (1997)] that approximates the likelihood with a product of appropriate conditional distributions. This yields a joint distribution, but not a process; hence, spatial interpolation is somewhat cumbersome. While promising, it is yet to be methodically explored for non-Gaussian and multivariate likelihoods such as ours. Yet another approach, known as "covariance tapering," considers compactly supported correlation functions [Furrer, Genton and Nychka (2006); Gneiting (2002)] that yield sparse correlation matrices. Efficient sparse solvers can then be devised for kriging and variance estimation, but these tapered functions may limit the scope of the models; also, full likelihood-based inference still requires determinant computations that may not be easily available. Recently Rue, Martino and Chopin (2009) propose a promising `INLA` (Integrated Nested Laplace Approximation) algorithm as an alternative to MCMC that delivers fast posterior approximations. The method's efficiency depends upon a Gaussian Markov random field approximation and may not be ideal with several hyperparameters (e.g., the matrix $\mathbf{A}$) such as ours.

4.1. *Predictive process models.* In principle, we might have adapted any of the above approaches to reduce the dimensionality of our model. Seeking a more seamless transition to multivariate processes, however, we opt for a class of models that emerges from representations of the spatial process in a lower-dimensional subspace and easily adapt to multivariate processes. These are often referred to as "low-rank" or "reduced-rank" spatial models and have been explored in different contexts [Lin et al. (2000); Stein (2007, 2008); Cressie and Johannesson (2008); Banerjee et al. (2008); Crainiceanu, Diggle and Rowlingson (2008)]. Many of these methods are variants of the so-called "subset of regressors" methods used in Gaussian process regressions for large data sets in machine learning [e.g., Wahba (1990); Rasmussen and Williams (2006) and references therein]. The idea here is to consider a set of locations, or "knots," say, $\mathcal{S}^* = \{\mathbf{s}_1^*, \ldots, \mathbf{s}_{n^*}^*\}$, where the number of knots is much smaller than the number of observed locations, and to represent the spatial process realizations over $\mathcal{S}$ in terms of the realizations over the smaller set of knots. Specifically, we define $\tilde{\mathbf{w}}(\mathbf{s}) = \mathcal{C}(\mathbf{s};\boldsymbol{\theta})'\Sigma_{\mathbf{w}}^*(\boldsymbol{\theta})^{-1}\mathbf{w}^*$, where $\mathcal{C}(\mathbf{s};\boldsymbol{\theta})'$ is the $p \times n^*p$ block matrix with $\mathbf{C_w}(\mathbf{s}, \mathbf{s}_j^*)$ being the $j$th block, $\Sigma_{\mathbf{w}}^*(\boldsymbol{\theta})$ is the $n^*p \times n^*p$ block matrix whose $(i, j)$th block is $\mathbf{C_w}(\mathbf{s}_i^*, \mathbf{s}_j^*)$ and $\mathbf{w}^* = (\mathbf{w}(\mathbf{s}_1^*)', \ldots, \mathbf{w}(\mathbf{s}_n^*)')'$ denotes a realization of the process $\mathbf{w}(\mathbf{s})$ over $\mathcal{S}^*$. Banerjee et al. (2008) call $\tilde{\mathbf{w}}(\mathbf{s})$ the *predictive process* derived from the *parent process* $\mathbf{w}(\mathbf{s})$. The appeal behind this formulation is that every spatial process (parent) model produces a corresponding predictive process model. In our subsequent models, we will assume a stationary cross-covariance for the parent process, that is, $\mathbf{C_w}(\mathbf{s}_1-\mathbf{s}_2)=\mathbf{A}\mathbf{C_v}(\mathbf{s}_1-\mathbf{s}_2)\mathbf{A}'$ and, in particular, $\mathbf{C_w}(\mathbf{s},\mathbf{s})=\mathbf{C_w}(\mathbf{0})=\mathbf{A}\mathbf{A}'$; the predictive process will still be nonstationary.

The dimension reduction is seen immediately. In fitting the predictive process counterparts of (3.1), the $np\times 1$ vector $\mathbf{w}=(\mathbf{w}(\mathbf{s}_1)',\dots,\mathbf{w}(\mathbf{s}_n)')'$ is replaced by $\mathbf{Z}(\boldsymbol{\theta})\mathbf{w}^*$, where $\mathbf{Z}(\boldsymbol{\theta})=\mathcal{C}(\boldsymbol{\theta})'\Sigma^*(\boldsymbol{\theta})^{-1}$ with $\mathcal{C}(\boldsymbol{\theta})'$ being an $np\times n^*p$ block matrix with $\mathbf{C_w}(\mathbf{s}_i,\mathbf{s}_j^*)$ as its $(i,j)$th block. Therefore, $\mathbf{Z}(\boldsymbol{\theta})$ is $np\times n^*p$ and we work with an $n^*p$-dimensional joint distribution only. Evidently, the parent model in (3.1) is different from its predictive process counterpart. Hence, though we introduce the same set of parameters in both models, they will not be identical. This approach leads to a different parameterization from that of low-rank smoothing splines [e.g., Lin et al. (2000); Kamman and Wand (2003); Crainiceanu, Diggle and Rowlingson (2008)], but yields the same joint marginal distribution for process realizations.

Stein (2007, 2008) undertakes an exploration of a subset of regressors methods, pointing out its pitfalls for data with fine-scale variation—a consequence, perhaps, of the fact that information for the covariances tends to concentrate at shorter lags. Indeed, the predictive process tends to oversmooth and also underestimates the spatial variance component. For multivariate processes, this follows from the following inequality:

$$\begin{aligned}\operatorname{var}\{\mathbf{w}(\mathbf{s})\}-\operatorname{var}\{\tilde{\mathbf{w}}(\mathbf{s})\}&=\mathbf{C_w}(\mathbf{s},\mathbf{s})-\mathcal{C}(\mathbf{s},\boldsymbol{\theta})'\Sigma^*(\boldsymbol{\theta})^{-1}\mathcal{C}(\mathbf{s},\boldsymbol{\theta})\\&=\operatorname{var}\{\mathbf{w}(\mathbf{s})|\mathbf{w}^*\}\succeq 0,\end{aligned}$$

where $\operatorname{var}\{\cdot\}$ denotes the variance–covariance matrix and $\succeq 0$ indicates nonnegative definiteness. Equality holds only when the knots coincide with the spatial locations, whence the predictive process realizations coincide with those from the parent process. The univariate analogue of the above argument shows that $\operatorname{var}(\tilde{w}_k(\mathbf{s}))\le\operatorname{var}(w_k(\mathbf{s}))$ for each element $k=1,\dots,p$.

A remedy for the bias is to add a spatially-varying noise to form $\tilde{\mathbf{w}}_{\tilde{\varepsilon}}(\mathbf{s})=\tilde{\mathbf{w}}(\mathbf{s})+\tilde{\varepsilon}(\mathbf{s})$, where $\tilde{\varepsilon}(\mathbf{s})\overset{\text{ind}}{\sim}N(\mathbf{0},\mathbf{C_w}(\mathbf{s},\mathbf{s})-\mathcal{C}(\mathbf{s},\boldsymbol{\theta})'\Sigma^*(\boldsymbol{\theta})^{-1}\mathcal{C}(\mathbf{s},\boldsymbol{\theta}))$. This "correction" yields $\operatorname{var}\{\tilde{\mathbf{w}}_{\tilde{\varepsilon}}(\mathbf{s})\}=\operatorname{var}\{\mathbf{w}(\mathbf{s})\}$ as desired and $E\{\tilde{\mathbf{w}}_{\tilde{\varepsilon}}(\mathbf{s})|\mathbf{w}^*\}=\tilde{\mathbf{w}}(\mathbf{s})$, so $\tilde{\mathbf{w}}_{\tilde{\varepsilon}}(\mathbf{s})$ inherits the attractive approximation properties of $\tilde{\mathbf{w}}(\mathbf{s})$. Also, no new parameters are introduced, ensuring identifiability, and the computational benefits are retained since $\tilde{\varepsilon}=(\tilde{\varepsilon}(\mathbf{s}_1)',\dots,\tilde{\varepsilon}(\mathbf{s}_n)')'$ has a block-diagonal variance–covariance matrix. In typical geostatistical models for continuous outcomes, $\tilde{\varepsilon}(\mathbf{s})$ is referred to as the "nugget" and is used to capture measurement error or micro-scale variability. Notice that such "nugget" effects do not arise naturally in generalized linear models, such as ours, and should not be interpreted as such. In fact, we do not introduce any "new" variance parameter for the nugget. Rather, $\tilde{\varepsilon}(\mathbf{s})$ has a very special variance structure that adjusts for the bias in the spatial variance, diminishes the excessive smoothness of $\tilde{\mathbf{w}}(\mathbf{s})$ and, in our experience, considerably improves model-fit and robustness to fine-scale variation.

We conclude this section with some brief remarks on knot selection. With a fairly even distribution of data locations, one possibility is to select knots on a uniform grid overlaid on the domain. A design-based approach that minimizes a spatially averaged predictive variance criterion [e.g., Zhu and Stein (2006); Diggle and Lophaven (2006)] can be used. With irregular locations, however, we may encounter substantial areas of sparse observations where placing would amount to "wastage," possibly leading to inflated variance estimates and slower convergence. More practical space-covering designs [e.g., Royle and Nychka (1998)] can yield a representative collection of knots that better cover the domain. We can also apply other popular clustering algorithms such as `k-means` or *partitioning around medoids* algorithms [e.g., Kaufman and Rousseeuw (1990)]. Implementations of these algorithms are available in `R` packages such as `fields` and `cluster` and have been used in spline-based low-rank kriging models [Ruppert, Wand and Carroll (2003)].

4.2. *Model fitting details.* Let $\tilde{\mathbf{w}}_{\tilde{\varepsilon}} = (\tilde{\mathbf{w}}_{\tilde{\varepsilon}}(\mathbf{s}_1)', \ldots, \tilde{\mathbf{w}}_{\tilde{\varepsilon}}(\mathbf{s}_n)')'$ denote the realizations from the noise-added predictive process over the set of observed locations in $\mathcal{S}$. This follows an $np \times 1$ multivariate normal distribution $\tilde{\mathbf{w}}_{\tilde{\varepsilon}} \sim N(\mathbf{0}, \Sigma_{\tilde{\mathbf{w}}_{\tilde{\varepsilon}}}(\boldsymbol{\theta}))$, $\Sigma_{\tilde{\mathbf{w}}_{\tilde{\varepsilon}}}(\boldsymbol{\theta}) = \Sigma_{\tilde{\varepsilon}}(\boldsymbol{\theta}) + \mathcal{C}(\boldsymbol{\theta})'\Sigma^*(\boldsymbol{\theta})^{-1}\mathcal{C}(\boldsymbol{\theta})$, and $\Sigma_{\tilde{\varepsilon}}(\boldsymbol{\theta})$ is a block-diagonal matrix whose $i$th diagonal is given by $\mathbf{C}_{\mathbf{w}}(\mathbf{s}_i, \mathbf{s}_i) - \mathcal{C}(\mathbf{s}_i; \boldsymbol{\theta})'\mathcal{C}^*(\boldsymbol{\theta})^{-1}\mathcal{C}(\mathbf{s}_i; \boldsymbol{\theta})$. Estimation could proceed with a multinomial likelihood that estimates the $\pi_j(\mathbf{s}_i)$'s in (3.1) from the posterior

$$p(\{\boldsymbol{\beta}_j, \boldsymbol{\theta}_j, \tilde{\mathbf{w}}_{\tilde{\varepsilon}_j}\}|\mathbf{Y}) \propto \prod_{j=1}^{J-1}\{p(\boldsymbol{\beta}_j)p(\boldsymbol{\theta}_j)p(\tilde{\mathbf{w}}_{\tilde{\varepsilon}_j}|\boldsymbol{\theta}_j)\} \times \left\{\prod_{i=1}^{n}\prod_{j=1}^{J}\pi_j(\mathbf{s}_i)^{Y_j(\mathbf{s}_i)}\right\},$$

where $\mathbf{Y}$ is the vector of observed outcomes. Alternatively, we find it more convenient to compute the posterior distributions for $J-1$ logistic regression models [see Agresti (2002), Chapter 7; Begg and Gray (1984)] and use the posterior samples to obtain the posterior distribution of the classifiers in (3.1). Now each posterior distribution is given by

$$\begin{aligned} p(\Omega_j|\mathbf{Y}) &\propto p(\boldsymbol{\beta}_j)p(\boldsymbol{\theta}_j)p(\tilde{\mathbf{w}}_{\tilde{\varepsilon}_j}|\boldsymbol{\theta}_j) \\ &\quad \times \prod_{i=1}^{n} p_j(\mathbf{s}_i)^{Y_j(\mathbf{s}_i)}(1 - p_j(\mathbf{s}_i))^{1-Y_j(\mathbf{s}_i)}, \end{aligned} \tag{4.1}$$

where $\Omega_j = \{\boldsymbol{\beta}_j, \tilde{\mathbf{w}}_{\tilde{\varepsilon}_j}, \boldsymbol{\theta}_j\}$, $\text{logit}\{p_j(\mathbf{s}_i)\} = \mathbf{x}_j(\mathbf{s}_i)'\boldsymbol{\beta}_j + \mathbf{z}_j(\mathbf{s}_i)'\tilde{\mathbf{w}}_{\tilde{\varepsilon}_j}(\mathbf{s}_i)$. Posterior estimation will proceed employing MCMC samples that will yield samples $\{\Omega_j^{(l)}\}_{l=1}^{L}$. This will then involve computing the expression in (4.1) featuring $\Sigma_{\tilde{\mathbf{w}}_{\tilde{\varepsilon}}}(\boldsymbol{\theta}_j)^{-1}$ for $j = 1, \ldots, J-1$ in each iteration of the MCMC. We benefit from the Sherman–Woodbury–Morrison (SWM) formula [see, e.g., Henderson and Searle (1981)] to compute $\Sigma_{\tilde{\mathbf{w}}_{\tilde{\varepsilon}}}(\boldsymbol{\theta}_j)^{-1}$ as

$$\begin{aligned} &\Sigma_{\tilde{\varepsilon}}(\boldsymbol{\theta}_j)^{-1} - \Sigma_{\tilde{\varepsilon}}(\boldsymbol{\theta}_j)^{-1}\mathcal{C}(\boldsymbol{\theta}_j)'[\Sigma^*(\boldsymbol{\theta}_j) + \mathcal{C}(\boldsymbol{\theta}_j)\Sigma_{\tilde{\varepsilon}}(\boldsymbol{\theta}_j)^{-1}\mathcal{C}(\boldsymbol{\theta}_j)']^{-1} \\ &\qquad \times \mathcal{C}(\boldsymbol{\theta}_j)\Sigma_{\tilde{\varepsilon}}(\boldsymbol{\theta}_j)^{-1}. \end{aligned}$$

Here $\Sigma_{\tilde{\varepsilon}}(\boldsymbol{\theta}_j)$ is block-diagonal, $\{\mathbf{C}_{\mathbf{w}}(\mathbf{s}_i, \mathbf{s}_i; \boldsymbol{\theta}_j) - \mathcal{C}(\mathbf{s}_i; \boldsymbol{\theta}_j)' \Sigma^*(\boldsymbol{\theta})^{-1} \mathcal{C}(\mathbf{s}_i; \boldsymbol{\theta}_j)\}$ being the $i$th block, while the other inversion in the second term is $n^*p \times n^*p$. Likelihood computations also require the determinant of $\Sigma_{\tilde{\mathbf{w}}_{\tilde{\varepsilon}}}(\boldsymbol{\theta}_j)$ given by $|\Sigma_{\tilde{\varepsilon}}(\boldsymbol{\theta}_j)||\Sigma^*(\boldsymbol{\theta}_j) + \mathcal{C}(\boldsymbol{\theta}_j)\Sigma_{\tilde{\varepsilon}}(\boldsymbol{\theta}_j)^{-1}\mathcal{C}(\boldsymbol{\theta}_j)'|/|\Sigma^*(\boldsymbol{\theta}_j)|$. Computational gains accrue because $|\Sigma_{\tilde{\varepsilon}}(\boldsymbol{\theta}_j)|$ is the product of the determinant of its block-diagonal submatrices, while the remaining two determinants are of order $n^*p$.

Spatial interpolation at a location $\mathbf{s}_0$ can be achieved by composition sampling: for each $\Omega_j^{(l)}$ drawn from the posterior, we first draw $\tilde{\mathbf{w}}_{\tilde{\varepsilon}_j}^{(l)}(\mathbf{s}_0) \sim p(\tilde{\mathbf{w}}_{\tilde{\varepsilon}_j}(\mathbf{s}_0)|\Omega_j^{(l)})$, which is multivariate normal with mean $\boldsymbol{\mu}_{\tilde{\mathbf{w}}_{\tilde{\varepsilon}_j}}(\mathbf{s}_0; \Omega_j) = \mathcal{C}_{\tilde{\mathbf{w}}_{\tilde{\varepsilon}_j}}(\mathbf{s}_0; \boldsymbol{\theta}_j)' \times \Sigma_{\tilde{\mathbf{w}}_{\tilde{\varepsilon}_j}}(\boldsymbol{\theta}_j)^{-1} \tilde{\mathbf{w}}_{\tilde{\varepsilon}_j}$ and variance $\Sigma_{\tilde{\mathbf{w}}_{\tilde{\varepsilon}_j}}(\mathbf{s}_0; \boldsymbol{\theta}_j) = \mathcal{C}_{\tilde{\mathbf{w}}_{\tilde{\varepsilon}_j}}(\mathbf{s}_0, \mathbf{s}_0; \boldsymbol{\theta}_j) - \mathcal{C}_{\tilde{\mathbf{w}}_{\tilde{\varepsilon}_j}}(\mathbf{s}_0; \boldsymbol{\theta}_j)' \times \Sigma_{\tilde{\mathbf{w}}_{\tilde{\varepsilon}_j}}(\boldsymbol{\theta}_j)^{-1} \mathcal{C}_{\tilde{\mathbf{w}}_{\tilde{\varepsilon}_j}}(\mathbf{s}_0; \boldsymbol{\theta}_j)$, where $\mathcal{C}_{\tilde{\mathbf{w}}_{\tilde{\varepsilon}_j}}(\mathbf{s}_0; \boldsymbol{\theta}_j)'$ is the $1 \times n$ block matrix with $i$th block given by $\mathcal{C}(\mathbf{s}_0, \boldsymbol{\theta}_j)' \Sigma^*(\boldsymbol{\theta}_j)^{-1} \mathcal{C}(\mathbf{s}_i; \boldsymbol{\theta}_j)$. The posterior distribution of the prediction probability $\pi_j(\mathbf{s}_0)$ is then directly obtained from (3.1) using the posterior samples for the $J-1$ models

$$\pi_j(\mathbf{s}_0)^{(l)} = \frac{\exp(\mathbf{x}_j(\mathbf{s}_0)'\boldsymbol{\beta}_j^{(l)} + \mathbf{z}_j(\mathbf{s}_0)'\tilde{\mathbf{w}}_{\tilde{\varepsilon}_j}^{(l)}(\mathbf{s}_0))}{1 + \sum_{k=1}^{J-1} \exp(\mathbf{x}_k(\mathbf{s}_0)'\boldsymbol{\beta}_k^{(l)} + \mathbf{z}_k(\mathbf{s}_0)'\tilde{\mathbf{w}}_{\tilde{\varepsilon}_k}^{(l)}(\mathbf{s}_0))}.$$

## 5. Data analysis.

5.1. *Model validation and benchmark comparisons.* The models detailed in Sections 3 and 4 were fit to the Michigan FIA data described in Section 2. Because our primary interest is in prediction of FTG, we compare the candidate models' ability to predict FTG for a set of 200 holdout (or validation) plots that were selected at random from the 5180 FIA plots. Prediction for a new (or holdout) plot is based on the prediction distribution, $\boldsymbol{\pi}(\mathbf{s}_0) = \{\pi_1(\mathbf{s}_0), \ldots, \pi_J(\mathbf{s}_0)\}$. We consider several different scoring rules to evaluate the predictive performance of the candidate models. A scoring rule provides a summary measure for evaluating a probabilistic prediction given the predictive distribution and the observed outcome. In our setting the scoring rule function is $S(\boldsymbol{\pi}, i)$, where $i$ is the index of the observed FTG. Given 200 holdout plots, $\{\mathbf{s}_{0q}\}_{q=1}^{200}$, we can calculate summary statistics of the scores, for example, the mean score is $\widehat{S} = \sum_{q=1}^{200} \frac{S(\boldsymbol{\pi}_q, i_q)}{200}$, where $\boldsymbol{\pi}_q = \boldsymbol{\pi}(\mathbf{s}_{0q})$. In fact, we can obtain the entire posterior distribution of the scoring rule [i.e., $S(\boldsymbol{\pi}^{(l)}, i)$, $l = 1, \ldots, L$] and report the posterior summaries. Gneiting and Raftery (2007) offer four scoring rules for prediction of categorical variables:

$$\text{Zero–one: } S(\boldsymbol{\pi}, i) = \begin{cases} 1, & \text{if } \pi_i = \max\{\pi_1, \ldots, \pi_J\}, \\ 0, & \text{if otherwise,} \end{cases}$$

$$\text{Quadratic: } S(\boldsymbol{\pi}, i) = 2\pi_i - \sum_{j=1}^{J} \pi_j^2 - 1,$$

$$\text{Spherical: } S(\boldsymbol{\pi}, i) = \frac{\pi_i}{(\sum_{j=1}^{J} \pi_j^2)^{1/2}},$$

$$\text{Logarithmic: } S(\boldsymbol{\pi}, i) = \log \pi_i.$$

Following definitions in Gneiting and Raftery (2007), all the noted scoring rules are strictly *proper* but for the zero–one, which is only proper. The zero–one scoring rule uses only a portion of available information, ignoring variability in the predictive distribution and returning either a zero or one. Similarly, the logarithmic scoring rule considers only one of the probabilities in the predictive distribution.

In addition to these four scoring rules, we examine classification confusion matrices, parameter estimates, and the models' ability to produce spatially consistent fitted and predicted distributions of FTGs.

We also compare our spatially-varying multinomial logistic regression models to common benchmark methods. Currently, $k$-nearest neighbor ($k$-NN) methods are among the most frequently applied for forest attribute mapping using NFI data [McRoberts, Nelson and Wendt (2002); Tomppo and Halme (2004)]. Using this method, the prediction probability that a new location will belong to the $j$th FTG is

$$\pi_j(\mathbf{s}_0) = \frac{1}{k} \sum_{l \overset{k}{\sim} 0} \delta_{j, y(\mathbf{s}_l)},$$

where we temporarily redefine $y(\mathbf{s})$ as the index (or label) belonging to one of the $J$ distinct FTG's and $\delta_{a,b}$ is the Dirac function ($\delta_{a,b} = 1$ if $a = b$, and 0 otherwise). The term $\frac{1}{k} \sum_{l \overset{k}{\sim} 0}$ records the proportion of the $j$th FTG in the set of $k$ nearest neighbors, where *nearness* is defined using a distance metric, $d(\cdot, \cdot)$, between the predictor variables. Here we consider two benchmark models: the first defines nearness as the Euclidean distance between geographic coordinates, $d(\mathbf{s}, \mathbf{s}')$, and the second calculates Euclidean distance between predictor vectors, $d(\mathbf{x}(\mathbf{s}), \mathbf{x}(\mathbf{s}'))$. The parameter $k$ is chosen by running the algorithm for a range of values, for example, $k \in 1, \ldots, 40$, then choosing the value that optimizes the specified objective function.

5.2. *Implementation specifics.* As detailed in Section 3, the candidate multinomial logistic regressions include nonspatial model, spatially-varying intercept and spatially-varying coefficients models. Maple/beech/birch is set as the regressions' baseline category due to its abundance (35.19% of observed plots) and pervasiveness [covering the entire domain, see Figure 2(b)].

For estimating predictive process models, we used 154, 200, and 254 knots over the domain [e.g., Figure 2(c) shows 200 knots]. We experimented with the clustering algorithms noted at the end of Section 3 and the infill designs suggested by Diggle and Lophaven (2006). Neither provided substantial changes in model parameter estimates or improvements in prediction. We note, however, that this lack of improvement is likely data specific and these knot selection approaches should be explored for each new analysis.

To complete the Bayesian specification, we assign priors to the models' parameters. As customary, we use a *flat* prior on all $\boldsymbol{\beta}$ parameters. For the univariate spatial process in the spatially-varying intercept model, priors must be specified for, $\sigma^2$, and the Matérn correlation function's range, $\phi$, and smoothness, $\nu$, parameters [i.e., $\boldsymbol{\theta} = \{\sigma^2, \phi, \nu\}$]. For each FTG model, we assume that $\sigma^2$ follows an $IG(2, 10)$ and $\phi$ follows a Uniform prior with support $U(5.22 \times 10^{-6}, 1)$, which is between 1 and 575,000 m when $\nu = 0.5$ (i.e., about 3/4 of the maximum intersite distance of 766.4 km). Again, with a shape value of 2, the $IG$ distribution has infinite variance and is centered on the scale value, which in this case is 10. The smoothness parameter was set to follow $U(0, 2)$. Then, as described in Section 3, the $n \times 1$ $\tilde{\mathbf{w}}_{\tilde{\varepsilon}}$ follows a $MVN(\mathbf{0}, \Sigma_{\tilde{\mathbf{w}}_{\tilde{\varepsilon}}}(\boldsymbol{\theta}))$.

For the spatially-varying coefficients model, we assumed $\mathbf{AA}'$ follows an $IW(p + 1, I_p)$ prior. We experimented with different diagonal $IW$'s scale matrices to assess this prior's influence on posterior distributions. The spatial range and smoothness parameter for each $\beta$ specific process again follow $U(5.22 \times 10^{-6}, 1)$ and $U(0, 2)$, respectively. The $np \times 1$ $\tilde{\mathbf{w}}_{\tilde{\varepsilon}}$ follows $MVN(\mathbf{0}, \Sigma_{\tilde{\mathbf{w}}_{\tilde{\varepsilon}}}(\boldsymbol{\theta}))$, with appropriate dimension adjustments to the mean vector and dispersion matrix. Details about the Metropolis sampling algorithm are provided in the Finley, Banerjee and McRoberts (2009) supplemental article.

For each model, three MCMC chains were run for 75,000 iterations. The sampler was coded in C++ and leveraged Intel's Math Kernel Library threaded BLAS and LAPACK routines for matrix computations. The code was run on a Linux workstation with two Intel Xeon quad core processors. The spatially-varying coefficients model was the most computationally challenging, with each chain of the 254 knot model taking ∼5 hours to complete. The CODA package in R (www.r-project.org) was used to diagnose convergence by monitoring mixing using Gelman–Rubin diagnostics and autocorrelations [see, e.g., Gelman et al. (2004), Section 11.6]. Acceptable convergence was diagnosed within 25,000 iterations and, therefore, 150,000 samples ($3 \times 50{,}000$) were retained for posterior analysis.

**6. Results and discussion.** Table 2 offers parameter estimates for the red/white/jack pine multinomial logistic regression candidate models [again, due to space limitations, parameter estimates for the remaining five FTGs are available in the Finley, Banerjee and McRoberts (2009) supplemental

TABLE 2

*Parameter credible intervals, $_{2.5\%}$ 50% $_{97.5\%}$ percentiles, for the red/white/jack pine FTG nonspatial and predictive process candidate model. Note, that for brevity, only the diagonal elements of* $\mathbf{C}$ *are provided for the space-varying coefficients model. All* $\phi$ *parameter values are scaled by* $10.0^5$ article]. For the red/white/jack pine FTG, as with the other FTGs, several of the predictors are significant at the 0.05 level. For the nonspatial and spatially-varying intercept models the parameters associated with SDI and SNOW were significant at the 0.05 level and the signs on these parameter estimates are consistent with the predictor surfaces in Figure 1(d, e) and observed red/white/jack plots in Figure 3(a). Specifically, the concentration of red/white/jack pine in the northcentral lower peninsula of Michigan correspond to areas of low snow depth and dry soils. The central most area of high red/white/jack pine in Figure 3(a) is predominantly composed of pure jack pine and red pine plantations. Recently established jack pine plantations came about as part of a state incentive program to reestablish habitat for the Kirtland's warbler (*Dendroica kirtlandii*), which is an endangered neotropical migratory bird. Similar connections between observed probability of FTG and predictor variables can be made for the other FTGs offered in the Finley, Banerjee and McRoberts (2009) supplemental article. Interpreting the significance of predictor variables' parameters for the spatially-varying coefficient models must be done in a spatial context. Rather than look at the aspatial $\boldsymbol{\beta}$ slope parameters, which for these models are simply the mean over the domain, interpretation should be based on $\tilde{\boldsymbol{\beta}}$, as depicted in Figure 6. However, it is important to interpret these parameter estimates with caution, recalling that presence/absence of an FTG is relative to the baseline category, not all other forested plots. If our focus was to better understand the relationship between FTGs and the environmental predictor variables, then we should use (2.1), and associated submodels, with the binary response of 1 if the FTG of interest is present and 0 otherwise, over all forested plots. Because our focus is on achieving high prediction accuracy, we do not dwell on interpreting the $\boldsymbol{\beta}$ or $\tilde{\boldsymbol{\beta}}$ and instead consider measures of predictive accuracy, model fit and model adequacy.

| | | | | Spatial predictive process | | | | | | | | | | | |
|---|---|---|---|---|---|---|---|---|---|---|---|---|---|---|---|
| | | | | Spatially-varying intercept | | | Spatially-varying coefficients | | | | | | | | |
| **Parameter** | **Nonspatial** | | | **200 knots** | | | **154 knots** | | | **200 knots** | | | **254 knots** | | |
| $\beta_{\text{RWJ; 0}}$ | −1.54 | −1.43 | −1.31 | −1.86 | −1.72 | −1.57 | −1.45 | −1.30 | −1.15 | −1.48 | −1.35 | −1.23 | −1.66 | −1.52 | −1.35 |
| $\beta_{\text{RWJ; PRECIP}}$ | −0.20 | −0.08 | 0.05 | −0.34 | −0.17 | 0.01 | −0.33 | −0.11 | 0.13 | −0.38 | −0.13 | 0.11 | 0.04 | 0.25 | 0.55 |
| $\beta_{\text{RWJ; TMAX}}$ | −0.11 | 0.12 | 0.33 | −0.08 | 0.24 | 0.46 | −0.26 | −0.02 | 0.22 | −0.29 | −0.05 | 0.23 | −0.52 | −0.28 | 0.06 |
| $\beta_{\text{RWJ; TMIN}}$ | −0.25 | −0.07 | 0.11 | −0.37 | −0.17 | 0.02 | −0.27 | −0.06 | 0.17 | −0.09 | 0.13 | 0.33 | −0.19 | 0.10 | 0.29 |
| $\beta_{\text{RWJ; SDI}}$ | −0.97 | −0.82 | −0.69 | −0.99 | −0.86 | −0.72 | −0.93 | −0.76 | −0.63 | −0.95 | −0.78 | −0.62 | −1.12 | −0.96 | −0.80 |
| $\beta_{\text{RWJ; SNOW}}$ | −0.50 | −0.34 | −0.20 | −0.60 | −0.39 | −0.14 | −0.71 | −0.45 | −0.26 | −0.49 | −0.23 | −0.11 | −0.92 | −0.68 | −0.48 |
| $\mathbf{C}_{\text{RWJ; 0}}$ and $\sigma^2$ | | – | | 3.63 | 4.46 | 5.73 | 4.62 | 6.04 | 7.69 | 3.96 | 5.14 | 6.89 | 5.11 | 5.86 | 6.92 |
| $\mathbf{C}_{\text{RWJ; PRECIP}}$ | | – | | | – | | 2.34 | 2.87 | 3.54 | 2.36 | 2.84 | 3.12 | 4.47 | 6.16 | 7.58 |
| $\mathbf{C}_{\text{RWJ; TMAX}}$ | | – | | | – | | 4.97 | 6.70 | 10.84 | 3.62 | 4.31 | 4.84 | 4.49 | 6.11 | 7.21 |
| $\mathbf{C}_{\text{RWJ; TMIN}}$ | | – | | | – | | 3.23 | 4.01 | 5.07 | 3.48 | 4.25 | 5.82 | 3.95 | 4.93 | 5.99 |
| $\mathbf{C}_{\text{RWJ; SDI}}$ | | – | | | – | | 3.65 | 6.24 | 7.58 | 3.31 | 4.03 | 5.21 | 3.57 | 4.19 | 4.82 |
| $\mathbf{C}_{\text{RWJ; SNOW}}$ | | – | | | – | | 4.00 | 4.79 | 5.55 | 3.19 | 3.86 | 5.06 | 3.61 | 4.52 | 5.85 |
| $\phi_{\text{RWJ; Intercept}}$ | | – | | 2.19 | 3.15 | 4.18 | 1.07 | 1.56 | 2.06 | 0.97 | 1.21 | 1.93 | 1.88 | 2.33 | 2.91 |
| $\phi_{\text{RWJ; PRECIP}}$ | | – | | | – | | 2.34 | 3.41 | 5.79 | 1.59 | 2.41 | 3.36 | 2.27 | 2.88 | 3.44 |
| $\phi_{\text{RWJ; TMAX}}$ | | – | | | – | | 0.63 | 0.69 | 0.82 | 2.44 | 3.76 | 5.27 | 1.30 | 2.16 | 3.18 |
| $\phi_{\text{RWJ; TMIN}}$ | | – | | | – | | 1.60 | 1.91 | 2.39 | 1.52 | 2.63 | 3.34 | 2.04 | 2.96 | 3.42 |
| $\phi_{\text{RWJ; SDI}}$ | | – | | | – | | 0.83 | 1.14 | 1.96 | 1.13 | 1.67 | 2.11 | 1.46 | 2.05 | 2.73 |
| $\phi_{\text{RWJ; SNOW}}$ | | – | | | – | | 1.94 | 2.56 | 3.23 | 1.96 | 2.40 | 2.72 | 1.43 | 1.87 | 3.14 |
| $\nu_{\text{RWJ; Intercept}}$ | | – | | 0.46 | 0.49 | 0.55 | 0.48 | 0.56 | 0.65 | 0.45 | 0.51 | 0.71 | 0.59 | 0.74 | 0.85 |
| $\nu_{\text{RWJ; PRECIP}}$ | | – | | | – | | 0.45 | 0.61 | 1.02 | 0.40 | 0.53 | 0.65 | 0.87 | 1.02 | 1.21 |
| $\nu_{\text{RWJ; TMAX}}$ | | – | | | – | | 0.38 | 0.47 | 0.57 | 0.77 | 1.01 | 1.38 | 0.52 | 0.77 | 0.90 |
| $\nu_{\text{RWJ; TMIN}}$ | | – | | | – | | 0.44 | 0.55 | 0.62 | 0.52 | 0.66 | 0.81 | 0.64 | 0.82 | 1.06 |
| $\nu_{\text{RWJ; SDI}}$ | | – | | | – | | 0.38 | 0.52 | 0.63 | 0.46 | 0.53 | 0.65 | 0.39 | 0.46 | 0.59 |
| $\nu_{\text{RWJ; SNOW}}$ | | – | | | – | | 0.68 | 0.81 | 0.94 | 0.49 | 0.55 | 0.73 | 0.41 | 0.60 | 0.74 |

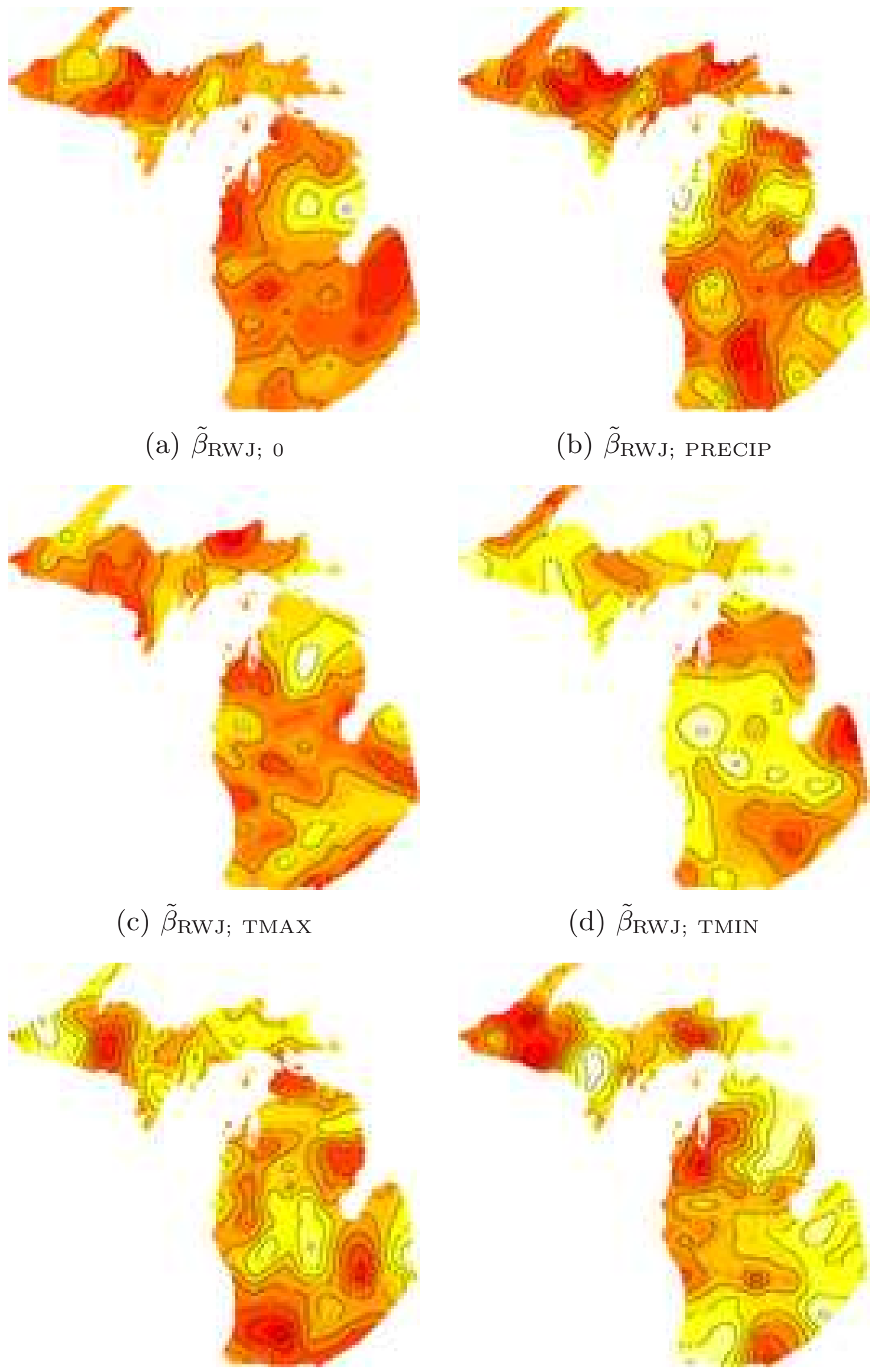

(a) $\tilde{\beta}_{\text{RWJ; 0}}$ (b) $\tilde{\beta}_{\text{RWJ; PRECIP}}$

(c) $\tilde{\beta}_{\text{RWJ; TMAX}}$ (d) $\tilde{\beta}_{\text{RWJ; TMIN}}$

(e) $\tilde{\beta}_{\text{RWJ; SDI}}$ (f) $\tilde{\beta}_{\text{RWJ; SNOW}}$

FIG. 6. *Interpolated surfaces of $\tilde{\beta}$ for the red/white/jack pine (RWJ) FTG. Lighter colors correspond to higher values.*

We now turn to the results of the 200 plot holdout set analysis. Table 3 displays scores for the benchmark and candidate models. The mean scores, over the 200 holdout plots, are reported for the benchmark models. For the multinomial models, fit using MCMC, the median and upper/lower 95% credible intervals for the scores' mean posterior distribution are reported. For the scoring rules, higher scores indicate superior predictive performance. That is, for the zero–one and spherical scoring rules, scores closer to 1 indicate greater predictive performance, whereas for the quadratic and logarithmic rules, scores closer to 0 suggest improved performance. The $k$-NN algorithm which measures nearness in geographic space, benchmark 1, provided the worst prediction on all scoring rules. Benchmark 2, which calculates nearness in predictor variable space, performed better than the nonspatial multinomial model (except for the logarithmic rule) and comparable to the space-varying intercept multinomial model for the quadratic and spherical rules. Noting the potential limitation of the zero–one and logarithmic scoring rules, this suggests that the easily implemented $k$-NN algorithm is a viable option for similar analyses, if second order properties of the resulting map products are not needed. Allowing the regression coefficients to vary spatially improved predictive performance over the single spatial random

TABLE 3

*Model comparisons using scoring rules for prediction of holdout set. Parameter credible intervals expressed as* $_{2.5\%}$ 50% $_{97.5\%}$ *percentiles*

| **Model** | | **Zero–one** | | | **Quadratic** | | | **Spherical** | | | **Logarithmic** | |
|---|---|---|---|---|---|---|---|---|---|---|---|---|
| Benchmark 1 | | 0.48 | | | −0.68 | | | 0.56 | | | −2.43 | |
| Benchmark 2 | | 0.53 | | | −0.61 | | | 0.62 | | | −1.55 | |
| Nonspatial multinomial | 0.49 | 0.51 | 0.53 | −0.64 | −0.63 | −0.63 | 0.60 | 0.60 | 0.60 | −1.30 | −1.29 | −1.28 |
| Spatially-varying intercept multinomial | | | | | | | | | | | | |
| 154 | 0.53 | 0.55 | 0.56 | −0.62 | −0.61 | −0.59 | 0.61 | 0.62 | 0.63 | −1.29 | −1.26 | −1.23 |
| 200 | 0.54 | 0.56 | 0.58 | −0.61 | −0.60 | −0.58 | 0.62 | 0.63 | 0.64 | −1.26 | −1.22 | −1.19 |
| 254 | 0.52 | 0.55 | 0.57 | −0.62 | −0.61 | −0.60 | 0.61 | 0.62 | 0.63 | −1.28 | −1.26 | −1.23 |
| Spatially-varying coefficients multinomial | | | | | | | | | | | | |
| 154 | 0.54 | 0.56 | 0.58 | −0.58 | −0.57 | −0.56 | 0.64 | 0.65 | 0.66 | −1.14 | −1.11 | −1.09 |
| 200 | 0.57 | 0.59 | 0.61 | −0.55 | −0.54 | −0.53 | 0.66 | 0.67 | 0.68 | −1.06 | −1.05 | −1.03 |
| 254 | 0.57 | 0.58 | 0.60 | −0.55 | −0.54 | −0.53 | 0.66 | 0.67 | 0.67 | −1.06 | −1.04 | −1.03 |

effect model, as reflected in the scores for the spatially-varying coefficients models. Increasing knot intensity beyond 200 did not improve predictive performance for the spatially-varying intercept and coefficients models.

The large difference in predictive ability between the nonspatial and spatial multinomial models reveals a requirement of the forest inventory sampling design and a limitation of the proposed models. Specifically, the observed locations must be dense enough to estimate the range of the spatial process associated with the intercept and, in the case of the spatially-varying regression model, predictor variables. If the data array is sufficiently sparse, or predictions are made beyond the range of the spatial influence, the prediction only learns from the predictor variables and cannot draw information from the proximity of the observed locations. For the analysis presented here, the FIA plot array is dense and predictions are made well within the support of the spatial range of the predictor variables.

Table 4 offers the confusion matrices for benchmark 2, nonspatial and 200 knot spatially-varying intercept and coefficients models. Here, prediction is based on the maximum FTG in the predictive distribution (equivalent to the zero–one scoring rule). Not surprisingly, these tables suggest substantial confusion within the conifer and deciduous FTG. For example, due to similar species composition, there is high misclassification between the maple/beech/birch and aspen/birch FTGs. Also, the relatively rare oak/pine FTG which exhibits a split of conifer and deciduous species was not correctly classified by any of the models. However, moving from the benchmark and nonspatial to the 200-knot spatially-varying intercept and then to the spatially-varying coefficients model does substantially improve prediction.

The validation analysis supports the use of the spatially-varying coefficients model. The use of this model is further corroborated by the presence of spatial dependence across the coefficients, which is summarized by the estimates of $\mathbf{C_w}(\mathbf{0})$, and coefficient specific $\phi$ and $\nu$ offered in Table 2. The estimated effective spatial ranges (i.e., the distance at which the spatial correlation drops to 0.05) and associated 95% credible intervals in kilometers for the red/white/jack pine FTG are 251 (173, 310), 124 (96, 173), 107 (85, 149), 127 (103, 203), 186 (154, 265), 135 (116, 155), for the intercept, PRECIP, TMAX, TMIN, SDI and SNOW, respectively. These long spatial ranges help support our initial simplifying assumption to combine the data layers into a common pixel resolution. The associated random spatial effect surfaces of $\tilde{\boldsymbol{\beta}}$ are again illustrated in Figure 6. Similar strong coefficient spatial dependence is seen in the other FTG models. The responsiveness of these spatially-varying coefficients to local trends in FTG presence/absence provides small errors between observed and fitted values as seen when comparing the left and right columns of Figures 3 and 4.

Table 4

*Confusion matrices from holdout set analysis for the benchmark 2, nonspatial multinomial, and 200 knot predictive process spatial multinomial models*

| | Predicted FTG | | | | | | | | | | | | | |
|---|---|---|---|---|---|---|---|---|---|---|---|---|---|---|
| **Observed FTG** | **WRJ** | **SF** | **OP** | **OH** | **EAC** | **MBB** | **AB** | **WRJ** | **SF** | **OP** | **OH** | **EAC** | **MBB** | **AB** |
| | Benchmark 2 | | | | | | | Nonspatial multinomial | | | | | | |
| WRJ | 9 | 0 | 0 | 6 | 0 | 7 | 3 | 0 | 1 | 0 | 11 | 0 | 13 | 0 |
| SF | 0 | 21 | 0 | 0 | 0 | 2 | 3 | 0 | 20 | 0 | 0 | 1 | 5 | 0 |
| OP | 1 | 0 | 0 | 2 | 1 | 2 | 0 | 1 | 0 | 0 | 2 | 1 | 2 | 0 |
| OH | 2 | 0 | 0 | 18 | 0 | 4 | 4 | 0 | 0 | 0 | 17 | 0 | 11 | 0 |
| EAC | 0 | 3 | 0 | 1 | 6 | 1 | 0 | 0 | 1 | 0 | 0 | 7 | 1 | 2 |
| MBB | 0 | 3 | 0 | 12 | 6 | 46 | 4 | 0 | 4 | 0 | 4 | 5 | 54 | 4 |
| AB | 1 | 4 | 0 | 7 | 2 | 14 | 5 | 0 | 4 | 0 | 4 | 1 | 21 | 3 |
| | Spatially-varying intercept multinomial | | | | | | | Spatially-varying coefficients multinomial | | | | | | |
| WRJ | 7 | 1 | 1 | 8 | 0 | 7 | 1 | 7 | 1 | 0 | 8 | 0 | 8 | 1 |
| SF | 1 | 19 | 0 | 0 | 1 | 0 | 5 | 0 | 23 | 0 | 1 | 1 | 0 | 1 |
| OP | 0 | 0 | 0 | 3 | 1 | 2 | 0 | 0 | 0 | 0 | 2 | 0 | 3 | 1 |
| OH | 2 | 0 | 2 | 17 | 0 | 4 | 3 | 0 | 0 | 1 | 21 | 0 | 3 | 3 |
| EAC | 0 | 1 | 0 | 2 | 6 | 0 | 2 | 0 | 0 | 0 | 1 | 6 | 2 | 2 |
| MBB | 0 | 4 | 0 | 7 | 3 | 54 | 3 | 2 | 2 | 0 | 6 | 4 | 52 | 5 |
| AB | 1 | 4 | 1 | 5 | 2 | 12 | 8 | 2 | 3 | 0 | 5 | 1 | 12 | 10 |

In practice, we can make a pixel-level prediction of the FTG predictive distribution wherever the predictor variable set is observed. Then, in a subsequent step using a Geographic Information System (GIS), users can remove or *mask* those pixels that were deemed nonforest from a separate prediction/classification exercise.

**7. Summary.** Analysis of large spatial domains is becoming more common, due, in part, to increased access to threaded mathematical libraries that can leverage the power of multi-processor computers and improvements in dimension reduction methods such as low-rank spatial processes. Only through a combination of these tools was the analysis presented here computationally feasible. With expanding domains of interest comes an increasing propensity for nonstationarity in the underlying spatial process. From a statistical validity standpoint, it is important that we define models that are equipped to deal with nonstationarity. From a utilitarian perspective, and as seen in our results, addressing nonstationarity can improve model fit and, more importantly, prediction.

Our central interest was to improve FTG prediction given a relatively dense array of forest inventory plots and a spatially complete set of environmental predictor variables. The results suggest that the multinomial logistic regression with spatially-varying coefficients is well suited to this objective. Further, following a Bayesian approach to model fitting provided estimates of the spatial parameters and access to posterior predictive probabilities at each new location. It would be difficult to estimate the spatial parameters a priori, and even if reasonable estimates could be made, the *plug-in* approach used in traditional methods can provide falsely precise estimates of predicted FTG probabilities. This could, in turn, negatively impact end-users sensitivity analyses.

Finally, while the current approach apparently meets our stated objectives, it does not account for underlying structured dependence across categories. Investigating and accounting for such dependence can be crucial in understanding the relationship between the probability of FTG occurrence and environmental variables, for example, how the spatial patterns of FTG's joint probabilities will shift with changing temperature or water regimes resulting from climate change. One approach is to consider spatial versions of the multinomial probit models [see, e.g., McCulloch, Polson and Rossi (2000)]. This and related issues are of great scientific interest and constructing models to elucidate these relationships will guide our future research effort.

**Acknowledgments.** The authors thank the referees, the Associate Editor and the Editor for valuable comments and suggestions. The authors also thank Dr. Huiyan Sang at the Department of Statistics Texas A&M University for several useful discussions regarding the predictive process.

## SUPPLEMENTARY MATERIAL

**Description of MCMC sampling algorithm and supplementary results.** (DOI: 10.1214/09-AOAS250SUPP; .pdf). Here we provide a description of the Metropolis scheme used to fit the candidate models. Parameter estimates for the FTGs are also presented.

A. O. Finley
Departments of Forestry and Geography
Michigan State University
East Lansing, Michigan
USA
E-mail: finleya@msu.edu

S. Banerjee
Division of Biostatistics
School of Public Health
University of Minnesota
Minneapolis, Minnesota
USA

R. E. McRoberts
Northern Research Station
USDA Forest Service
Saint Paul, Minnesota
USA